\documentclass[trackchanges]{aastex62}
\listofchanges
\graphicspath{{./}{Figure/}}
\usepackage[single=true]{acro}
\usepackage{lineno}
\usepackage{xspace}
\usepackage{upgreek}
\usepackage{color}
\usepackage{nicefrac}
\newcommand{\gtlike}{\tt{gtlike}}

\newcommand{\fer}{{\itshape Fermi} LAT\xspace}
\newcommand{\ve}{\varepsilon}
\newcommand{\sub}[1]{_{\mbox{\tiny #1}}}
\newcommand{\SF}{seven}
\DeclareAcronym{pwn}{
  short = PWN ,
  long  = pulsar wind nebula ,
  short-plural = e ,
  long-plural = e ,
  class = astro ,
  first-style = default
}

\DeclareAcronym{mhd}{
  short = MHD ,
  long  = magnetohydrodynamics ,
  class = hydro ,
  first-style = default
}
\DeclareAcronym{ic}{
  short = IC ,
  long  = inverse Compton ,
  class = astro ,
  first-style = default
}

\shorttitle{Crab small flares}
\shortauthors{Arakawa et al.}
\begin{document}

\title{Detection of Small Flares from the Crab Nebula with \textit{Fermi}-LAT}

\correspondingauthor{Masanori Arakawa}
\email{m.arakawa@rikkyo.ac.jp}

\author[0000-0002-0786-7307]{Masanori Arakawa}
\affil{Department of Physics, Rikkyo University, 3-34-1 Nishi-Ikebukuro, Toshima-ku, Tokyo 171-8501, Japan}
\affil{Astrophysical Big Bang Laboratory, RIKEN, Saitama 351-0198, Japan}

\author{Masaaki Hayashida}
\affiliation{Department of Physics, Rikkyo University, 3-34-1 Nishi-Ikebukuro, Toshima-ku, Tokyo 171-8501, Japan}
\affiliation{Graduate School of Artificial Intelligence and Science, Rikkyo University, 3-34-1 Nishi-Ikebukuro, Toshima-ku, Tokyo 171-8501, Japan}
\affiliation{Galaxies, Inc., 1-1-11 Minami-Ikebukuro, Toshima-ku, Tokyo, 171-0022, Japan}

\author[0000-0002-7576-7869]{Dmitry Khangulyan}
\affiliation{Department of Physics, Rikkyo University, 3-34-1 Nishi-Ikebukuro, Toshima-ku, Tokyo 171-8501, Japan}

\author{Yasunobu Uchiyama}
\affiliation{Department of Physics, Rikkyo University, 3-34-1 Nishi-Ikebukuro, Toshima-ku, Tokyo 171-8501, Japan}
\affiliation{Graduate School of Artificial Intelligence and Science, Rikkyo University, 3-34-1 Nishi-Ikebukuro, Toshima-ku, Tokyo 171-8501, Japan}
\affiliation{Galaxies, Inc., 1-1-11 Minami-Ikebukuro, Toshima-ku, Tokyo, 171-0022, Japan}



\begin{abstract}
%% 250 word limit
  Gamma radiation from the Crab \ac{pwn} shows significant variability at $\sim100$~MeV energies, recently revealed with spaceborne gamma-ray telescopes.
  Here we report the results of a systematic search for gamma-ray flares  using a 7.4-year data set acquired with the {\itshape Fermi} Large Area Telescope. Analyzing the off-pulse phases of the Crab pulsar,
  we found {\SF} previously unreported low-intensity flares (``small flares").
  The ``small flares'' originate from the variable synchrotron component of the Crab \ac{pwn} and show clearly different features from the steady component of the Crab \ac{pwn} emission. They are characterized by larger fluxes and harder photon indices, similar to previously reported flares.
  These flares show day-scale time variability and imply a strong magnetic field of
  $B_{\rm min}\approx 1~\mathrm{mG}$ at the site of the gamma-ray production.
  This result seems to be inconsistent with the typical values revealed with modeling of the non-thermal emission from the nebula.
  The detection of the ``small flares'' gives a hint of production of gamma rays above $100$~MeV in a
  part of the nebula with properties which are different from the main emitters, e.g., due to bulk relativistic motion.
\acresetall

\end{abstract}
%\keywords{editorials, notices ---
%miscellaneous --- catalogs --- surveys}
\section{Introduction} \label{sec:intro}
\acresetall

The Crab pulsar and its \ac{pwn}  are among the most studied objects in the Galaxy.
The central pulsar has a period of 33\,ms and large spin-down power, $\dot{E}\sub{SD}\simeq5\times10^{38}~\mathrm{erg\,s^{-1}}$ \citep{Hester2008}.
Almost all of $\dot{E}\sub{SD}$ is carried away by an ultra-relativistic wind mainly composed of electron-positron pairs (hereafter electrons).
The electrons are accelerated and randomized at the termination shock, which is located $\sim$0.14~pc from the pulsar \citep{Weisskopf2000}.
Downstream of the termination shock, the interaction of the accelerated electrons with the magnetic and photon fields results in the production of broadband non-thermal radiation spanning radio to multi-TeV energies.

While the synchrotron emission provides the dominant contribution from radio to GeV energies, the emission produced through the \ac{ic} scattering is responsible for the gamma rays detected above $\sim$1 GeV.
The broadband spectrum of the Crab \ac{pwn} is consistent with a 1-D hybrid kinetic--\ac{mhd} approach, in which radiative models account for the advective transport of particles, radiative and adiabatic cooling, and spatial distributions of magnetic and photon fields in the \ac{pwn} \citep{Kennel1984b,Atoyan1996}.
These models allow one to reproduce the broadband spectral energy distribution, assuming a very low pulsar wind magnetization, $\sigma$.
This requirement resulted in the formulation of the so-called $\sigma$ problem.
It represents a mismatch of the wind magnetization at the light cylinder, which is expected to be very high $(\sigma\gg1)$, and the one inferred with 1D MHD models for \acp{pwn} $(\sigma \sim10^{-3})$.

In addition, 1D MHD models do not allow one to study the morphology seen in the center part of the Crab PWN, namely the jet, torus and wisps \citep[e.g.,][]{Hester2008}.
2D MHD models, which adopt an anisotropic pulsar wind, can consistently explain the jet-torus morphology, as shown in theoretical studies \citep{2002AstL...28..373B,2002MNRAS.329L..34L} and in 2D numerical simulations \citep{2004MNRAS.349..779K,2006A&A...453..621D}.
The properties of wisps, e.g., their emergence and velocity, can also be explained by such 2D MHD simulations \citep[see, e.g., ][]{2008A&A...485..337V,Camus2009, 2015MNRAS.449.3149O}.
Finally, the 3D MHD simulations successfully reproduce the morphological structures and provide a possible solution for the $\sigma$ problem due to a significant magnetic field dispersion inside the PWN \citep{2014MNRAS.438..278P, 2016JPlPh..82f6301O}.

Recently, spaceborne gamma-ray telescopes ({\textit{AGILE} and \fer}) revealed that \(\sim100\)~MeV  emission from the Crab PWN displays day-scale variability \citep{2011Sci...331..736T, Crab_flare_1st}.
This short variability implies that this emission is produced through the synchrotron channel by electrons with \(\sim\)PeV energies. The cooling time for the synchrotron emission in days is determined by
\begin{equation}
    t\sub{SYN}=100 \left(\frac{\ve}{100~\mathrm{MeV}}\right)^{-1/2}\left(\frac{B}{100~\mathrm{\upmu G}}\right)^{-3/2}\rm\, days\,,
\end{equation}
where $\ve$ and $B$ are the mean energy of emitted photons and the magnetic field, respectively.
The rapid variability of flares requires a very strong magnetic field, $\gtrsim 1~\mathrm{mG}$  \citep[e.g.,][]{flare_review}, which significantly exceeds the average magnetic field in the nebula, $B=100-300~\rm \mu G$ \citep[see][and references therein]{2020MNRAS.491.3217K}.
The recent 3D MHD simulation indicates such a strong magnetic field can exist at the base of the plume \citep[see, e.g.,][]{2014MNRAS.438..278P, 2016JPlPh..82f6301O}.

Another important argument for production of flares under very special conditions comes from the peak energy.  The
spectral energy distribution of the 2011 April and 2013 March flares are characterized by cut-off energies of $375\pm 26$ MeV and
$484^{+409}_{-166}$ MeV, respectively  \citep{Rolf2012,Mayer2013}.  Thus, the spectra extend beyond the maximum synchrotron peak energy, $\sim$ 236 MeV, attainable in the ideal-\ac{mhd} configurations  \citep[the synchrotron burn-off limit, see, e.g.,][]{2002PhRvD..66b3005A,Arons2012}. The synchrotron peak frequency can exceed this limit if (i) particles are accelerated in the non-\ac{mhd} regime  \citep[e.g., via magnetic reconnection, see][]{2012ApJ...754L..33C}, (ii) if  the emission is produced in a relativistically moving outflow  \citep{2004MNRAS.349..779K},  or (iii) if small scale magnetic turbulence is present   \citep{2013ApJ...774...61K}. All these possibilities\footnote{The magnetic turbulence on the scale required for the jitter regime might be suppressed by the Landau damping in PWNe \citep[see, e.g.,][]{2019arXiv190507975H}, thus this possibility seem to be less feasible  \citep[see, however,][]{2012MNRAS.421L..67B}.} imply production of flares under very special circumstances, which differ strongly from the typical conditions expected in the Crab \ac{pwn}. The physical conditions at the production site of the stationary/slow-varying synchrotron gamma-ray emission are less constrained, and this radiation component can be generated under the same conditions as the dominant optical-to-X-ray emission, which in particular implies a magnetic field of \(\sim0.1\rm\,mG\) \citep[see, e.g.,][]{Kennel1984b,Aharonian1998,2014MNRAS.438..278P}.

The synchrotron gamma rays show variability on all resolvable time scales \citep{Rolf2012}.
The current detection of flares is based on automated processing of the LAT data   \citep{Atwood2009}. Such analysis includes both the radiation from the Crab pulsar and \ac{pwn}.

The phase-averaged photon flux above 100\,MeV is dominated by the Crab pulsar emission with $\sim 2\times10^{-6}~\mathrm{photon\,cm^{-2}\,s^{-1}}$ \citep[e.g.,][]{Crab_pulsar2010}.
Analysis using only off-pulse phase data allows us to study the \ac{pwn} synchrotron gamma-ray emission free from the Crab pulsar emission.
While an off-pulse analysis loses some photon statistics due to reduced exposure time, it can minimize systematic uncertainties caused by estimations of the Crab pulsar flux.
It helps in investigating even small variations of the \ac{pwn} synchrotron emission more reliably.
Detections of small-flux-scale flares might give a clue to deepen our understanding of the physical phenomena responsible for the flaring emission.
Here, we present a systematic search for gamma-ray flares using 7.4 years of data obtained by the Large Area Telescope (LAT) on board {\textit{Fermi}}, specifically analyzing the off-pulse phases of the Crab pulsar.
This paper is organized as follows.
In Section \ref{section2}, we describe the analysis procedure and the results of the long-term (7.4 years) and shorter term (30 days, 5 days, and 1.5 days) analysis.
In Section \ref{section3}, we discuss the physical interpretation of the ``small flares.'' The conclusion is in Section \ref{section4}.
\newpage
\section{Observation and data analysis} \label{section2}
\subsection{Observation and data reduction}\label{section2.1}
 {\it Fermi}-LAT is a $e^{\pm}$ pair-production detector covering the 20\,MeV--$>$300\,GeV energy range.  LAT is composed of a converter/tracker made of layers of Si strips and Tungsten to convert and then measure the direction of incident photons,
 a calorimeter made of CsI scintillator to determine the energy of gamma rays,
 and an anti-coincidence detector to reject background charged particles \citep{Atwood2009}.
 LAT has a large effective area ($>$ 8200 $\mathrm{cm^{2}}$) and a wide field of view ($\sim$ 2.4 str).
 The point spread function (PSF) of the LAT, which becomes better with increasing energy, is about 0.9 deg (at 1 GeV) and 0.1 deg (at 10 GeV) .

We analyzed the data from MJD 54686 (2008 August 8) to 57349 (2015 November 11) in the energy range between 100 MeV and 500 GeV.
We adopted the low-energy threshold of 100 MeV to reduce systematic errors, especially originating from energy dispersion, although our choice leads to smaller photon statistics than previous studies \citep{Rolf2012,Mayer2013}, which used a lower energy threshold of 70\,MeV.
The data analysis was performed using the Science Tools package (v11r05p3) distributed by the Fermi Science Support Center (FSSC) following the standard procedure\footnote{\url{http://fermi.gsfc.nasa.gov/ssc/data/analysis/}} with the P8R2$\_$SOURCE$\_$V6 instrument response functions.
Spectral parameters were estimated by the maximum likelihood using {\gtlike} implemented in the Science Tools.
We examined the detection significance of gamma-ray signals from sources by means of the test statistic (TS) based on the likelihood ratio test  \citep{Mattox}.

Our analyses are composed of two parts: the longer-term (7.4 years) scale for the baseline state, and the shorter-term (30 days, 5 days, and 1.5 days) scale for flare states.
For the longer-term analysis, the events were extracted within a 21.2 $\times$ 21.2 degree region of interest (RoI) centered on the Crab PWN position (RA: 83.6331 deg, Dec: 22.0199 deg).
After the standard quality cut (DATA$\_$QUAL$>$0$\&\&$(LAT$\_$CONFIG==1), the events with zenith angles above 90 deg were excluded to reduce gamma-ray events from the Earth limb.
The data when the Crab PWN was within 5 deg of the Sun were also excluded.
The data were analyzed by the binned maximum likelihood method.
The background model includes sources within 18 degrees of the Crab PWN as listed in the Preliminary LAT 8-year Point Source List (FL8Y)\footnote{\url{https://fermi.gsfc.nasa.gov/ssc/data/access/lat/fl8y/}}.
The radius of 18 degrees completely encloses the RoI.
Suspicious point sources, FL8Y J0535.9+2205 and FL8Y J0531.1+2200, which are located near the Crab PWN ($<$ 0.8 deg), were excluded\footnote{The fourth Fermi Large Area Telescope source catalog  \citep{4fgl} has just recently been published and does not include those two sources.}.
The supernova remnants IC 443 and S 147 are included as spatially extended templates.
Both normalizations and spectral parameters of the sources which are located within 5 degrees of the Crab PWN are set free while the parameters of all other sources are fixed at the FL8Y values.
The Galactic diffuse emission model ``gll$\_$iem$\_$v06.fits'' and the isotropic diffuse emission one ``iso$\_$P8R2$\_$SOURCE$\_$V6$\_$v06.txt'' are both included in the background model.
A normalization factor and photon index for the Galactic diffuse emission model and a normalization factor for the isotropic diffuse emission model are set free.
Sources with TS$<1$ were removed after the first iteration of the maximum likelihood fit, and then we fitted the data again.
We excluded the known flare periods as reported in  \cite{Mayer_thesis} and  \cite{Rudy2015} to determine the baseline state in our long-term analysis.
We defined the flare periods as two weeks before or after the peak flare times.
The flare periods which were excluded in our long-term analysis are summarized in Table \ref{reported_flare}.
In this paper, we refer those flares as ``reported flares''.
\begin{table}[b]
\caption{Periods of “reported flares”. These periods were excluded in our long-term analysis to determine the baseline state of the source.}
\label{reported_flare}
  \begin{center}
  \begin{tabular}{l|c|c}
    Flare name& MJD& Reference\\\hline\hline
    2009 February&54855 - 54883& \cite{Mayer_thesis}\\
    2010 September&55446 - 55474& \cite{Mayer_thesis}\\
    2011 April&55653 - 55681& \cite{Mayer_thesis}\\
    2012 July&56098 - 56126& \cite{Mayer_thesis}\\
    2013 March&56343 - 56371& \cite{Mayer_thesis}\\
    2013 October\tablenotemark{\rm \dag}&56568 - 56608& \cite{Rudy2015}\\
    2014 August\tablenotemark{\rm \dag}&56869 - 56902& \cite{Rudy2015}\\\hline
  \end{tabular}
  \tablenotetext{\rm \dag}{The light curve shows a two-peak structure (\cite{Rudy2015}).}
   \end{center}
\end{table}

For the shorter-term analysis, the events were extracted within a 15-degree
acceptance cone centered on the location of the Crab PWN,
and the gamma-ray fluxes and spectra were determined by the unbinned maximum likelihood method.
The background model is the same one as used in the long-term analysis, but parameters are fixed by the results from our long-term analysis, except for the isotropic diffuse emission, whose normalization remains free.

{\it Fermi}-LAT cannot spatially distinguish gamma rays originating from the Crab pulsar and PWN due to its large PSF.
Thus, the off-pulse window of the Crab pulsar is needed to obtain an accurate spectrum of the Crab PWN.
For this purpose, we used the \textsc{Tempo2} package\footnote{\url{http://www.atnf.csiro.au/research/pulsar/tempo2/index.php?n=Main.HomePage}}  \citep{tempo2} for phase gating analysis. The ephemeris data were prepared following the methods outlined in \cite{Kerr2015}.
The duration of the ephemeris data of the Crab pulsar is between MJD 54686 and 57349, and the phase interval 0.56--0.88 was chosen to  suppress effects from the pulsar.
All subsequent analysis  was performed using the off-pulse data of the Crab pulsar following \citet{Crab_flare_1st}.

\subsection{Spectral model}
The spectrum of the Crab PWN has two components in the LAT energy band.
One is a synchrotron component, which has a soft spectrum, and the other is an IC component which dominates above 1 GeV.
We assume a power-law spectrum (PL) for the synchrotron component and logparabola (LP) spectrum for the IC component:
\begin{equation}\label{crab_stationary}
\frac{dN}{dE_{\gamma}}=N\sub{0,SYN}\left(\frac{E_{\gamma}}{100~\mathrm{MeV}}\right)^{-\Gamma\sub{0,SYN}} + N\sub{0,IC}\left(\frac{E_{\gamma}}{1~\mathrm{GeV}}\right)^{-(\alpha_{0}+\beta_{0}\ln{(E_{\gamma}/1~\mathrm{GeV})})}
\end{equation}
We performed a maximum likelihood analysis of data between MJD 54686 and 57349 excluding the ``reported flares'' as defined in Sec \ref{section2.1}, and obtained the spectral values $\Gamma\sub{0,SYN}=4.27\pm 0.08$, $\alpha_{0}=1.50\pm0.04$, $\beta_{0}=0.05\pm0.01$.
The photon flux above 100 MeV is $F\sub{ph,SYN} = (6.31\pm0.23)\times10^{-7}~\mathrm{photon~cm^{-2}~s^{-1}}$ for the synchrotron component and $F\sub{ph,IC} = (1.09\pm0.08)\times10^{-7}~\mathrm{photon~cm^{-2}~s^{-1}}$ for the IC component, whereas the energy flux is $F\sub{e,SYN} = (1.45\pm0.06)\times10^{-10}~\mathrm{erg~cm^{-2}~s^{-1}}$ and $F\sub{e,IC} = (5.46\pm0.28)\times10^{-10}~\mathrm{erg~cm^{-2}~s^{-1}}$.
In the following sections, we refer to these values as the baseline values.

\begin{figure}[hbp]
\begin{center}
\includegraphics[width=95mm]{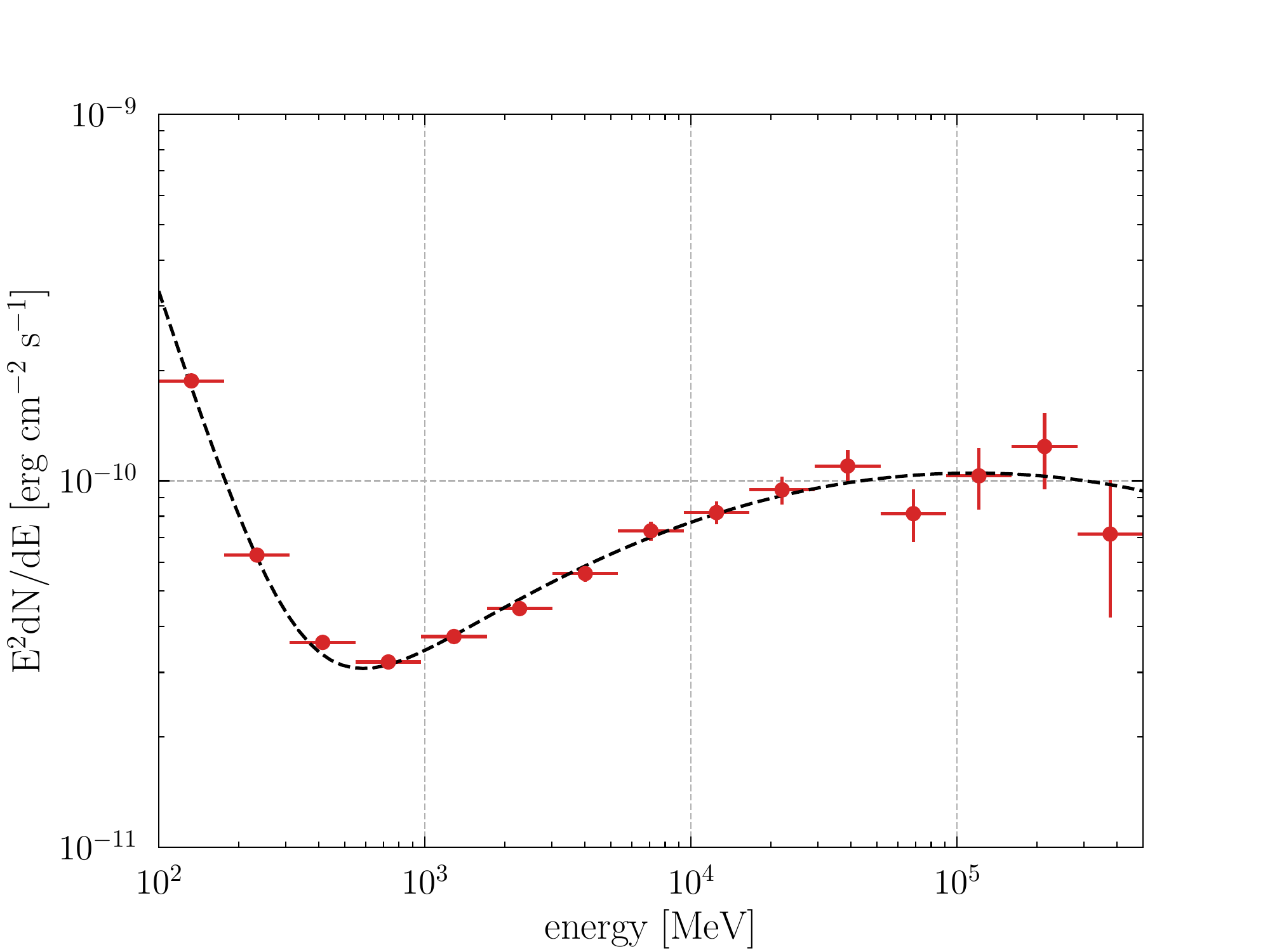}
\caption{Spectral energy distribution of the Crab PWN averaged over 7.4 years of \textit{Fermi}-LAT observations,  excluding the ``reported flares'' listed in Table \ref{reported_flare}. Only the data of the off-pulse phase (phase interval: [0.56-0.88]) were used.
The dashed black line represents the best-fitted model of the Crab PWN (synchrotron component $+$ Inverse Compton component) described by Eq. (\ref{crab_stationary}).}
\label{sed_pwn}
\end{center}
\end{figure}

Figure \ref{sed_pwn} presents the baseline spectral energy distribution of the Crab PWN.
The spectral points were obtained by dividing the 100 MeV to 500 GeV range into 15 logarithmically spaced energy bins.
%%%%%%%%%%%%%%%%%%%%%%%%%%%%%%%%%%%%%%%%%%%%%%%%%%%%%%%%%%%%%%%%%%%%%%%%%%%%%%%%%%%%%%%%%%%%%%%%%%%%%%%%
%%%%%%%%%%%%%%%%%%%%%%%%%%%%%%%%%%%%%%%%%%%%%%%%%%%%%%%%%%%%%%%%%%%%%%%%%%%%%%%%%%%%%%%%%%%%%%%%%%%%%%%%
\subsection{Temporal analysis}
\subsubsection{Day-scale analysis}\label{day_scale_ana}
\begin{figure*}[b]
\begin{center}
\includegraphics[width=125mm]{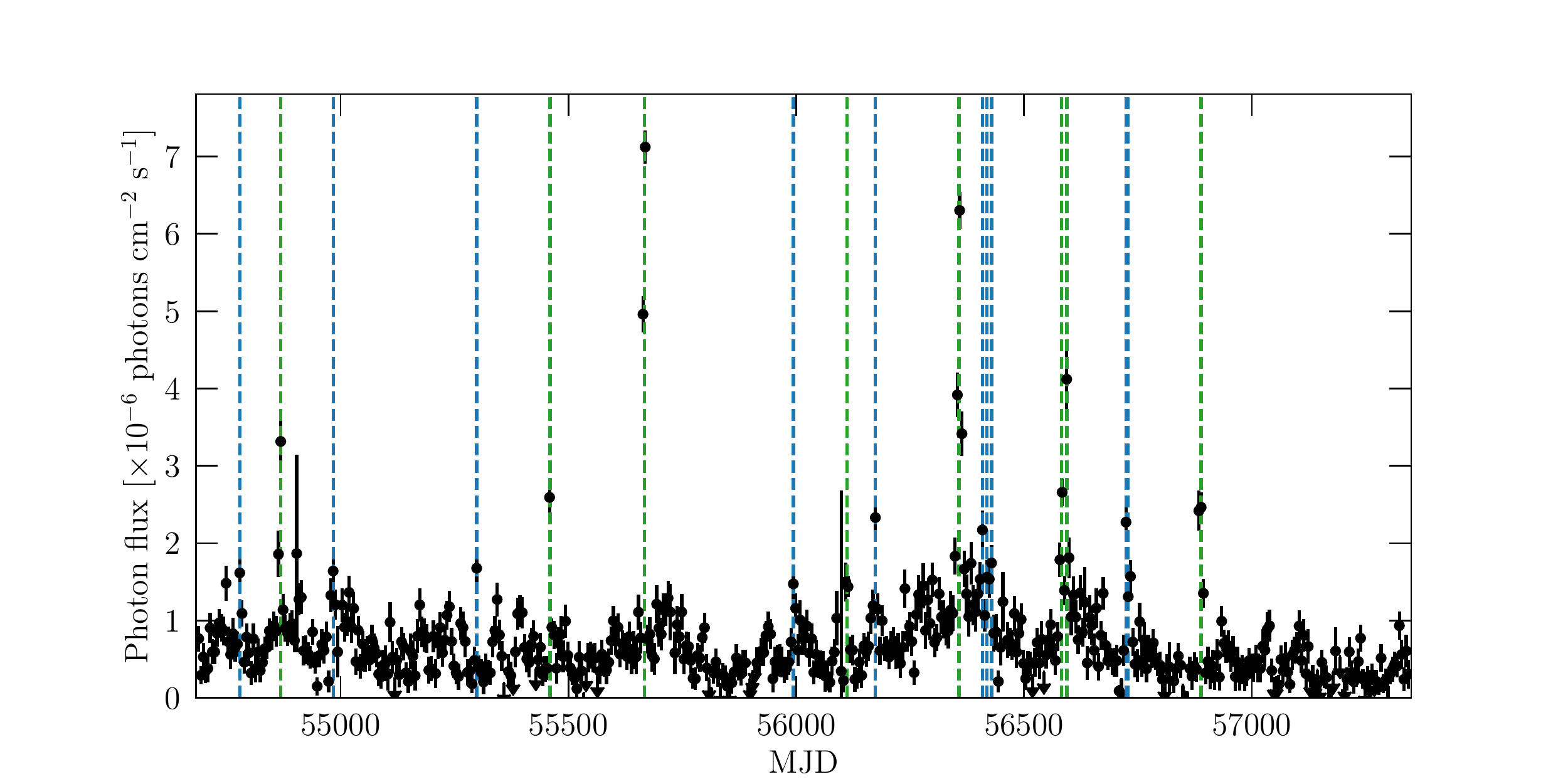}
\caption{5-day gamma-ray light curve (100 MeV -- 500 GeV integral photon flux) of the Crab synchrotron component from 2008 August  to 2015 November.
The green lines show the times of the reported flares listed in Table~\ref{reported_flare}.
The center times of the small flares from this work are indicated in blue lines as listed in Table \ref{flare_time}.}
\label{lc_5day}
\end{center}
\end{figure*}

We derived a 5-day binned light curves (LC) of the synchrotron and the IC components for the energy range 100\,MeV--500\,GeV,  based on  Eq. (\ref{crab_stationary}). We then used the $\chi^2$ test to examine variability.
The 5-day scale was chosen to match observed flare durations of a few days to $\sim$1 week \citep{flare_review}.
We treated the normalization of the synchrotron component, the IC component and the isotropic diffuse emission as free parameters while the others were fixed by the baseline values.
For the $\chi^2$ test, we excluded the data bins with TS$<2$ and the bins overlapping with the  reported flares  listed in Table \ref{reported_flare}.
The $\chi^2$ is defined as follows:
\begin{equation}
    \chi^2\sub{SYN/IC} = \sum_{i}\frac{(F\sub{i, SYN/IC}-F\sub{base, SYN/IC})^2}{F\sub{i, err,SYN/IC}^2}
\end{equation}
where $F\sub{i, SYN/IC}$ and $F\sub{i,err, SYN/IC}$ are the values of the flux and error of the synchrotron and IC component in each LC bin, respectively.
The derived $\chi^2\sub{SYN/IC}/$(d.o.f) are $\chi^2\sub{SYN}=1230.46/432$ and $\chi^2\sub{IC}=511.07/459$.
The synchrotron component is highly variable, while the IC component is less variable.
It has been reported that the IC component does not show any variability by previous LAT analysis  \citep{Crab_flare_1st, Rolf2012, Mayer2013} and by ground-based imaging air Cherenkov telescope observations  \citep{HESS_flare, VERITAS_flare, MAGIC_flare}.
Thus in the following analysis, we assume the spectral parameters of the IC component ($N\sub{0,IC},~\alpha_{0} \mathrm{~and~}\beta_{0}$) to be fixed at the baseline values.
The 5-day LC of the synchrotron component appears in Figure \ref{lc_5day}.

To identify  flare activity of the synchrotron component of the Crab PWN emission, we modeled the emission as a superposition of three components; steady synchrotron, IC and additional flare components, similar to   \cite{Crab_flare_1st}.
The flare component is modeled by a PL with free normalization and photon index parameters, while the steady synchrotron and IC components are fixed to the baseline values.
We define ``small flares'' as those whose TS for the flare component exceeds 29.
This choice corresponds to a significance of $\sim 5\sigma$ with 2 degrees of freedom (pre-trial) or $\sim 3.7\sigma$ considering trials for the 525 LC bins.
There is not a unique method to define ``a Crab flare'' since the Crab PWN emission is variable on all observed time scales \citep{Rolf2012}.
The criterion using TS is affected by differences of exposures among individual time bins and might overlook a high flux state if the exposure is rather short at that bin.
On the other hand, we can probe significances of the variation even on a small-flux scale with the TS value.
The times (MJD) of the centers of the peak 5-day LC bins and the TS of detected ``small flares'' are summarized in Table {\ref{flare_time}}.
These center times of small flares are also shown in Figure \ref{lc_5day} as blue lines.
All ``reported flares'' listed in Table \ref{reported_flare} satisfy the criterion of TS$>$29.
\begin{table}[tb]
\caption{Detected ``small flares'' in the 5-day binned light curve.}
\label{flare_time}
  %\begin{tabular}{|l|l|l||l||l|l|l||l||l|} \hline
  \begin{center}
  \begin{tabular}{c|c|c}
    Name& Bin midpoint (MJD)& TS (significance\tablenotemark{ i})\\\hline\hline
    small flare 1&54779&32.6 (4.1$\sigma$)\\
    small flare 2\tablenotemark{ii}&54984&32.4 (4.1$\sigma$)\\
    small flare 3&55299&34.4 (4.3$\sigma$)\\
    small flare 4\tablenotemark{iii}&55994&37.3 (4.6$\sigma$)\\
    small flare 5&56174&78.2 (7.8$\sigma$)\\
    small flare 6 (a), (b), (c)&56409, 56419, 56429&59.4, 30.2, 30.9 (6.5$\sigma$, 3.8$\sigma$, 3.9$\sigma$)\\
    small flare 7 (a), (b)\tablenotemark{iv}& 56724, 56734&93.4, 66.0 (8.7$\sigma$, 7.1$\sigma$)\\\hline
  \end{tabular}
  \tablenotetext{i}{Corresponding significance with 2 degrees of freedom with 525 trials.}
   \tablenotetext{ii}{Indicated as the minor flare in  \citet{Striani2013}.}
    \tablenotetext{iii}{Indicated as the ``wave'' in  \citet{Striani2013}.}
    \tablenotetext{iv}{ATel \#5971  \citep{Atel2014MAr}.}
   \end{center}
\end{table}

\begin{figure*}[tbp]
\begin{center}
\includegraphics[width=160mm]{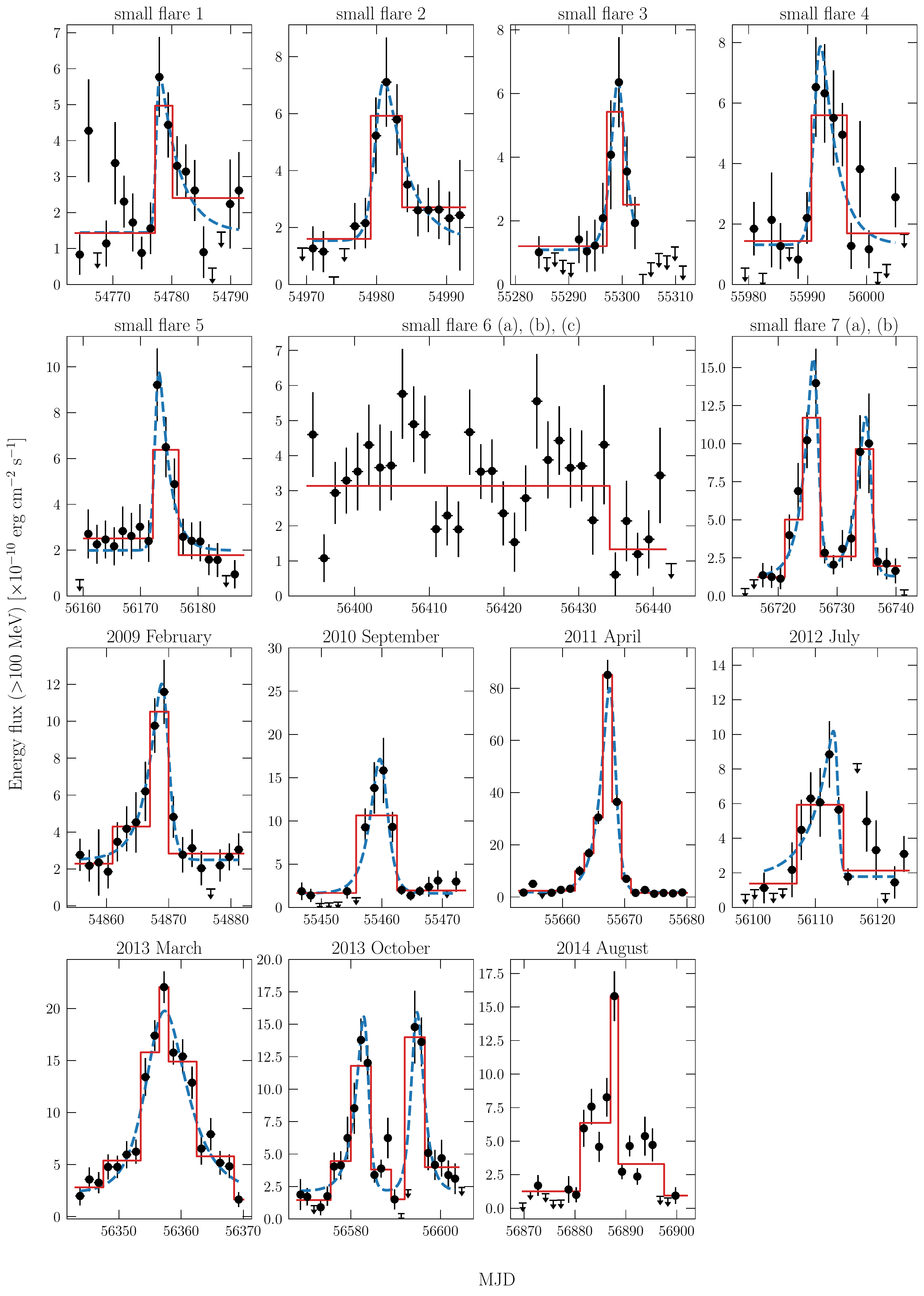}
\caption{1.5-day binned gamma-ray light curve (100 MeV -- 500 GeV) of the Crab synchrotron component during each ``small flare'' and ``reported flare''.
 The vertical error bars in data points represent 1$\sigma$ statistical errors. The down arrows indicate 95$\%$ confidence level upper limits.
 The dashed blue lines and red solid lines represent the best fitted time profiles defined by Eq. (\ref{lc_profile}) and the Bayesian Blocks, respectively.}

 \label{1.5days LC}
 \end{center}
\end{figure*}

In order to analyze detailed structures of ``small flares'' and ``reported flares'' we made 1.5-day binned LCs for one month for the small flare 1--5 and 7, and 50 days for the small flare 6 centered at the small-flare times listed in Table~\ref{flare_time} and the over the durations of ``reported flares'' listed in Table~\ref{reported_flare}.
The 1.5-day time bin was chosen rather than a 1-day bin to retain significant detections of synchrotron \ac{pwn} emission even at the baseline state while still allowing the resolution of flare structure.
In the same manner as the 5-day binned LC in Figure \ref{lc_5day}, the Crab PWN is modeled by Eq. (\ref{crab_stationary}) and $N\sub{0,SYN},~\Gamma\sub{0,SYN}$, with the normalization of the isotropic diffuse emission set free.
Figure \ref{1.5days LC} represents the 1.5-day binned LCs in the energy range 100 MeV to 500 GeV during ``small-flare'' and ``reported-flare'' periods.
%\newpage
%\noindent
The LCs are fitted by the following function to characterize the time profiles of both ``small flares” and ``reported flares'':
%\textcolor{red}{exclude ``In order to ... the following function:''}
\begin{equation}\label{lc_profile}
    %F(t)=F_\mathrm{b}+\frac{F_0}{e^{-(t-t_0)/\tau\sub{rise}} + e^{(t-t_0)/\tau\sub{decay}}}
    F(t)=F\sub{b} + \sum_i^{N}\frac{F\sub{i,0}}{e^{-(t-t\sub{i,0})/\tau\sub{i, rise}}+e^{-(t-t_{i,0})/\tau\sub{i,decay}}}
\end{equation}
where $N$ is the number of flares in each flare window, $F_\mathrm{b}$ is an assumed constant level underlying a flare, $F\sub{i,0}$ is the amplitude of a flare, $t\sub{i,0}$ describes approximately a peak time (it corresponds to the actual maximum only for symmetric flares), while $\tau\sub{i,rise}$ and $\tau\sub{i,decay}$ describe the characteristic rise and decay times.
Note that $F_\mathrm{b}$ does not represents the global baseline level, $F\sub{e,SYN}$, but a local synchrotron level, which reflexes the variability of the synchrotron component (see Figure~\ref{lc_5day}).
The time of the maximum of a flare ($\tau\sub{peak}$) can be described using parameters in Eq. (\ref{lc_profile}) as:
\begin{equation}\label{peak time}
    t\sub{peak}=t_0 +\frac{\tau\sub{rise}\tau\sub{decay}}{\tau\sub{rise}+\tau\sub{decay}}\ln\left(\frac{\tau\sub{decay}}{\tau\sub{rise}}\right)
\end{equation}
This formula is often used to characterize flare activities of blazars  \citep[e.g.,][]{hayashida2015}.
We applied the Bayesian Block (BB) algorithm \citep{2013ApJ...764..167S} to the 1.5-day binned LCs to determine the number of flares.
The BB procedure allows one to obtain the optimal piecewise representations of the LCs.
The calculated piecewise model provides local gamma-ray variabilities \citep[see, e.g.,][]{2019ApJ...877...39M}.
Since small flare 6 (a), (b) and (c) are not represented by piecewise models, we exclude these ``small flares" in the following analysis.
The rise time of the 2013 October flare is assumed to be 1 day because of lack of significant data points to determine the rise time.
The 2014 August flare is not fitted well by Eq. (\ref{lc_profile}), perhaps  because of statistical fluctuations or because the flare is a superposition of short and weak flares that cannot be resolved by the {\textit{Fermi}}-LAT.
Consequently, we do not estimate the characteristic time scales of the 2014 August flare.
The fitting results are summarized in Table~{\ref{flare_summary}}, and the best fitted profiles and the optimal piecewise models are overlaid in Figure \ref{1.5days LC} as blue dashed lines and red lines, respectively.
\begin{table*}[tbp]
\caption{Fitting results for the 1.5-day light curves}
\label{flare_summary}
  %\begin{tabular}{|l|l|l||l||l|l|l||l||l|} \hline
  \begin{center}
  \begin{tabular}{c|c|c|c|c|c}\hline\hline
    flare id& $F\sub{b}$  & $F_0$ & $\tau\sub{rise}$ & $\tau\sub{decay}$ & $t_{0}$\\
    &  [$\times10^{-10} ~\mathrm{erg}~\mathrm{cm^{-2}}~\mathrm{s^{-1}}$] & [$\times10^{-10} ~\mathrm{erg}~\mathrm{cm^{-2}}~\mathrm{s^{-1}}$] & [day] & [day] & [MJD]\\\hline
    small flare 1&1.4$~\pm~$0.2&5.6$~\pm~$2.0&0.3$~\pm~$0.4&3.3$~\pm~$1.7&54777.4$~\pm~$1.0\\
    small flare 2&1.5$~\pm~$0.4&9.0$~\pm~$2.6&0.7$~\pm~$0.5&3.2$~\pm~$1.3&54980.2$~\pm~$0.7\\
    small flare 3&1.1$~\pm~$0.4&10.6$~\pm~$2.8&1.1$~\pm~$1.0&1.3$~\pm~$0.9&55299.1$~\pm~$1.6\\
    small flare 4&1.3$~\pm~$0.4&10.6$~\pm~$3.1&0.6$~\pm~$0.4&2.7$~\pm~$1.1&55991.4$~\pm~$0.6\\
    small flare 5&2.0$~\pm~$0.3&12.5$~\pm~$4.2&0.4$~\pm~$0.2&1.8$~\pm~$0.8&56172.7$~\pm~$0.5\\
    small flare 7 (a)&1.3$~\pm~$0.5\tablenotemark{ \dag}&24.6$~\pm~$5.8&1.6$~\pm~$0.6&0.5$~\pm~$0.1&56726.4$~\pm~$0.4\\
    small flare 7 (b)&1.3$~\pm~$0.5\tablenotemark{ \dag}&18.2$~\pm~$6.3&1.7$~\pm~$0.7&0.6$~\pm~$0.3&56735.3$~\pm~$0.6\\\hline\hline
    2009 February&2.5$~\pm~$0.3&16.3$~\pm~$3.8&2.3$~\pm~$0.9&0.7$~\pm~$0.3&54869.5$~\pm~$0.6\\
    2010 September&1.6$~\pm~$0.3&30.0$~\pm~$6.6&1.7$~\pm~$0.6&1.0$~\pm~$0.3&55460.0$~\pm~$0.6\\
    2011 April&1.8$~\pm~$0.3&142.7$~\pm~$10.8&1.6$~\pm~$0.1&0.6$~\pm~$0.1&55668.0$~\pm~$0.1\\
    2012 July&1.8$~\pm~$0.6&11.2$~\pm~$3.5&3.3$~\pm~$1.4&0.3$~\pm~$1.2&56113.6$~\pm~$0.8\\
    2013 March&2.3$~\pm~$0.5&34.2$~\pm~$2.0&2.4$~\pm~$0.4&3.6$~\pm~$0.6&56356.8$~\pm~$0.6\\
    2013 October (a)&2.1$~\pm~$0.3\tablenotemark{\dag} & 23.5$~\pm~$3.6&2.1$~\pm~$0.5&0.7$~\pm~$0.2&56583.4$~\pm~$0.3\\
    2013 October (b)&2.1$~\pm~$0.3\tablenotemark{\dag}&27.0$~\pm~$4.8&1.0 (fixed)&1.5$~\pm~$0.5&56594.5$~\pm~$0.6\\\hline
  \end{tabular}
  \tablenotetext{\dag}{Fitted by the same $F\sub{b}$ because of the same flare window.}
   \end{center}
   \vspace{5mm}
\end{table*}

\begin{figure}[bp]
 \begin{minipage}{0.5\hsize}
  \begin{center}
   \includegraphics[width=95mm]{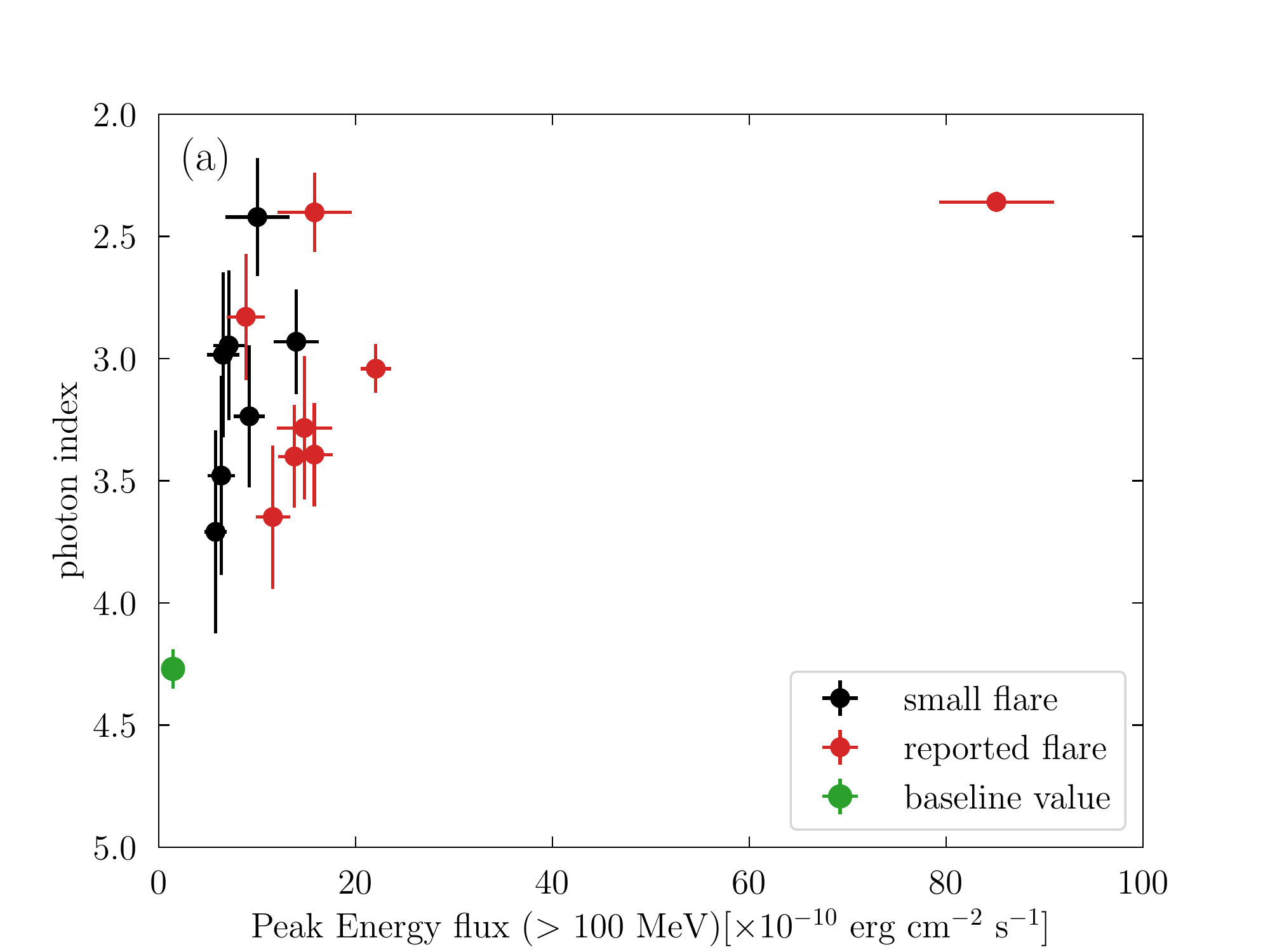}
  \end{center}
 \end{minipage}
 \begin{minipage}{0.5\hsize}
  \begin{center}
   \includegraphics[width=95mm]{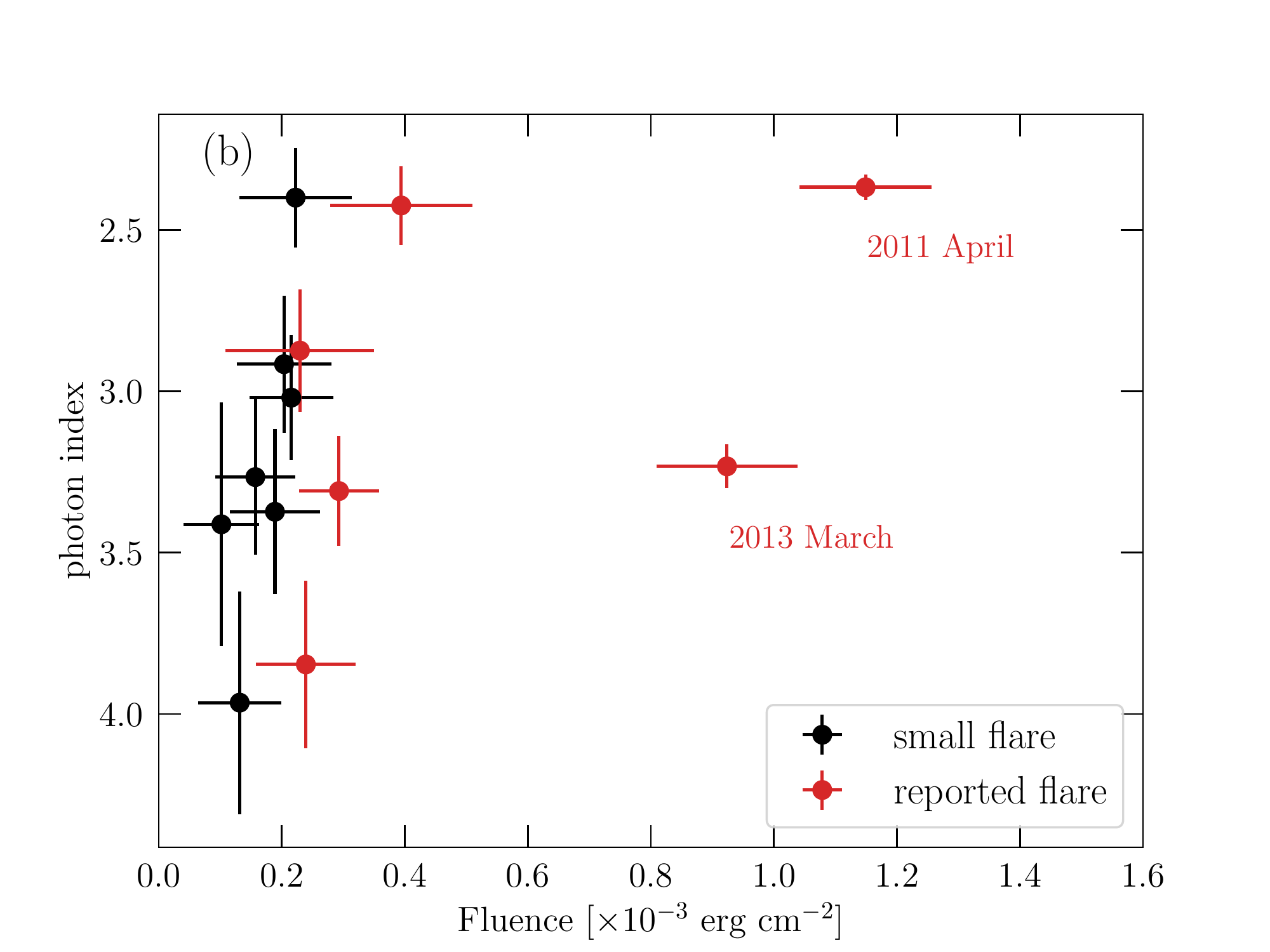}
  \end{center}
 \end{minipage}
 \caption{(a): Photon index vs. peak energy flux (100 MeV - 500 GeV) for the Crab flares.
 Black data points: ``small flares''.  Red points: ``reported flares''.
 Green data point:  baseline value.
 (b): Photon index vs. fluence for the Crab flares.
 The fluence is defined as integrated emission over a period between $t\sub{peak} -\tau\sub{rise}$  to $t\sub{peak} +\tau\sub{decay}$.
 Black data points:  ``small flares''. Red points: ``reported flares''.
 The 2014 August flare are not shown in this plot because they have complex time profiles
 (note that 2013 October (b) flare is excluded because the rise time is not determined).
 }
 \label{lc_corr}
\end{figure}
We compare energy fluxes and photon indices at the flare peaks in Figure \ref{lc_corr}-(a), and fluences and photon indices in Figure \ref{lc_corr}-(b) for individual flares. The flare peak is defined as the highest flux bin in each panel of Figure \ref{1.5days LC}, and the energy fluxes and photon indices were taken from the results of corresponding bins.
The flare fluences were defined with integrations over periods between between $t\sub{peak} -\tau\sub{rise} \leq t \leq t\sub{peak} +\tau\sub{decay}$, based on the best fitted time profile model as listed in Table \ref{flare_summary}.
The fluence and photon index were calculated with {\gtlike} from data resampled over those integration periods.
The fitting results are summarized in Table \ref{flare_spec}.
Figure \ref{lc_corr}-(a) indicates that the flares have higher energy fluxes and harder photon indices than the time-averaged values. This feature suggests both the ``small flares" and the ``reported flares" have different origins from the stationary emission.
The relation between the flux and photon index suggests a ``harder when brighter'' trend, which could imply electron spectra with higher cut-off energies or stronger magnetic fields for the flaring states.
In the fluence vs. photon index plot (Figure \ref{lc_corr}-(b)),
two flares clearly show high fluences.
One corresponds to the 2011 April flare, with the largest energy flux $\sim 85\times10^{-10}$ erg~cm$^{-2}$ s$^{-1}$,
and the other is the 2013 March flare ,with the second largest flux and longest flare duration (see Table \ref{flare_summary})\footnote{in the 2013 March flare, rapid variability on a $\sim$5-hour time scale has been reported using orbit-binned ($\sim$ 90 minutes) LC  \citep{Mayer2013}. Our analysis is based on 1.5-day binned LC and focuses on more global features of the flare.}.
Apart from those two flares, the ``reported flares'' and the ``small flares'' show similar results: the same range of photon index, and the harder/brighter correlation.
The rise time of the 2013 October flare (b) is not determined, therefore it is excluded in Figure \ref{lc_corr}-(b).

To examine possible spectral curvature, we applied not only a power-law model, but also a power law with an exponential cut-off model for the Crab synchrotron component. Five flares (small flare 7 (b), 2011 April flare, 2013 March flare, 2013 October flare (a) and October flare (b)) show significant curvature ($-2\Delta L >9$)\footnote{$-2\Delta L = -2\log(L0/L1)$, where $L0$ and $L1$ are the maximum likelihood estimated for the null (a simple power-law model) and alternative (a power law with an exponential cut-off model) hypothesis, respectively.} and those results are listed in Table \ref{flare_spec}.

\begin{table*}[tb]
\caption{Spectral fitting results of each flare}
\label{flare_spec}
  %\begin{tabular}{|l|l|l||l||l|l|l||l||l|} \hline
  \begin{center}
  \begin{tabular}{c|c|c|c|c|c}\hline\hline
    flare id&flare duration & averaged energy flux  & photon index & cut-off energy & $-2\Delta L$ \\
    &[day]&  [$\times10^{-10} ~\mathrm{erg}~\mathrm{cm^{-2}}~\mathrm{s^{-1}}$] &  & [MeV] &\\\hline
    small flare 1&3.5 $\pm$ 1.7&4.3$~\pm~$0.6& 4.0$~\pm~$0.3&--&--\\
    small flare 2&3.9 $\pm$ 1.4&5.6$~\pm~$0.8& 3.4$~\pm~$0.3&--&--\\
    small flare 3&2.4 $\pm$ 1.3&5.0$~\pm~$1.0& 3.4$~\pm~$0.4&--&--\\
    small flare 4&3.3 $\pm$ 1.1&7.1$~\pm~$1.2& 2.9$~\pm~$0.2&--&--\\
    small flare 5&2.2 $\pm$ 0.8&8.3$~\pm~$1.2& 3.3$~\pm~$0.2&--&--\\
    small flare 7 (a)&2.2$~\pm~$0.6&11.6 $\pm$  1.6& 3.0 $\pm$ 0.2&--&--\\
    small flare 7 (b)&2.3$~\pm~$0.8&11.4 $\pm$  2.6& 2.4 $\pm$ 0.2&--&--\\
    2009 February&3.0 $\pm$ 0.9&9.4$~\pm~$1.1& 3.8$~\pm~$0.3&--&--\\
    2010 September&2.6 $\pm$ 0.6&17.4$~\pm~$3.0& 2.4$~\pm~$0.1&--&--\\
    2011 April&2.2 $\pm$ 0.2&60.4$~\pm~$3.8& 2.37$~\pm~$0.04&--&--\\
    2012 July&3.6 $\pm$ 1.8&7.3$~\pm~$1.1& 2.9$~\pm~$0.2&--&--\\
    2013 March&6.0 $\pm$ 0.7&17.9$~\pm~$0.7& 3.2$~\pm~$0.1&--&--\\
    2013 October (a)&2.7$~\pm~$0.5&12.3 $\pm$  1.2& 3.3 $\pm$ 0.2&--&--\\
    2013 October (b)&2.5\tablenotemark{\dag}&15.6 $\pm$  1.8& 3.1 $\pm$ 0.2&--&--\\\hline
    smallflare 7 (b)&--&8.6$~\pm~$ 1.3&0.6$~\pm~$0.6&284$~\pm~$101&12.8\\
    2011 April&--&45.5$~\pm~$1.8& 1.6$~\pm~$0.1&688$~\pm~$115&71.2\\
    2013 March&--&17.0$~\pm~$0.6& 2.2$~\pm~$0.2&257$~\pm~$ 67&20.5\\
    2013 October (a)&--&11.7$~\pm~$ 1.1&1.4$~\pm~$0.8&135$~\pm~$68&9.1\\
    2013 October (b)&--&14.5$~\pm~$ 1.5&0.6$~\pm~$0.9&115$~\pm~$50&13.6\\\hline
  \end{tabular}
  \tablenotetext{\dag}{Rise time, $\tau\sub{rise}$ is fixed by 1 day.}
  \tablecomments{The upper section presents the fitting results using a power-law model. The lower section shows the results also using
  a power law with an exponential cut-off model for the flares in which significant curvature ($-2\Delta L>9$) was observed.
  $\Delta L$ presents the difference of the logarithm of the likelihood of the fit with respect to a single power-law fit.
  }
   \end{center}

\end{table*}

\clearpage
\subsubsection{Month-scale analysis}
The Crab PWN synchrotron emission in the gamma-ray band shows variability not only on a day scale but also on a month scale  \citep{Crab_flare_1st}.
One of the most prominent variable morphological features of the Crab PWN is known as the ``inner knot" \citep{Hester1995}.
The ``inner knot'' lies about 0.55--0.75 arcsec to the southeast of the Crab pulsar.
This structure is interpreted as Doppler-boosted emission from the downstream of the oblique termination shock.
The bulk of the synchrotron gamma rays may originate from the ``inner knot" according to MHD simulations \citep{Komissarov2011}.
 In addition, the ``inner knot" is a promising production site for the flares, as the Doppler boosting can relax the theoretical constraints imposed by the Crab flares, such as exceeding the maximum cut-off energies under the ideal-MHD configuration. In this case the gamma-ray flux level and location of the ``inner knot'' can change coherently.
%Although  \citet{Rudy2015} attempted to link the gamma-ray flare flux and the the separation distance between the Crab pulsar and ``inner knot" (knot-pulsar separation), they found no significant correlation.
\citet{Rudy2015} compared the gamma-ray flux and the knot-pulsar separation and found no significant correlation.
In their analysis, the contribution of the Crab pulsar emission was not excluded, making it difficult to measure the weaker flux variations of the Crab PWN.

\begin{figure*}[tb]
\begin{center}
\includegraphics[width=150mm]{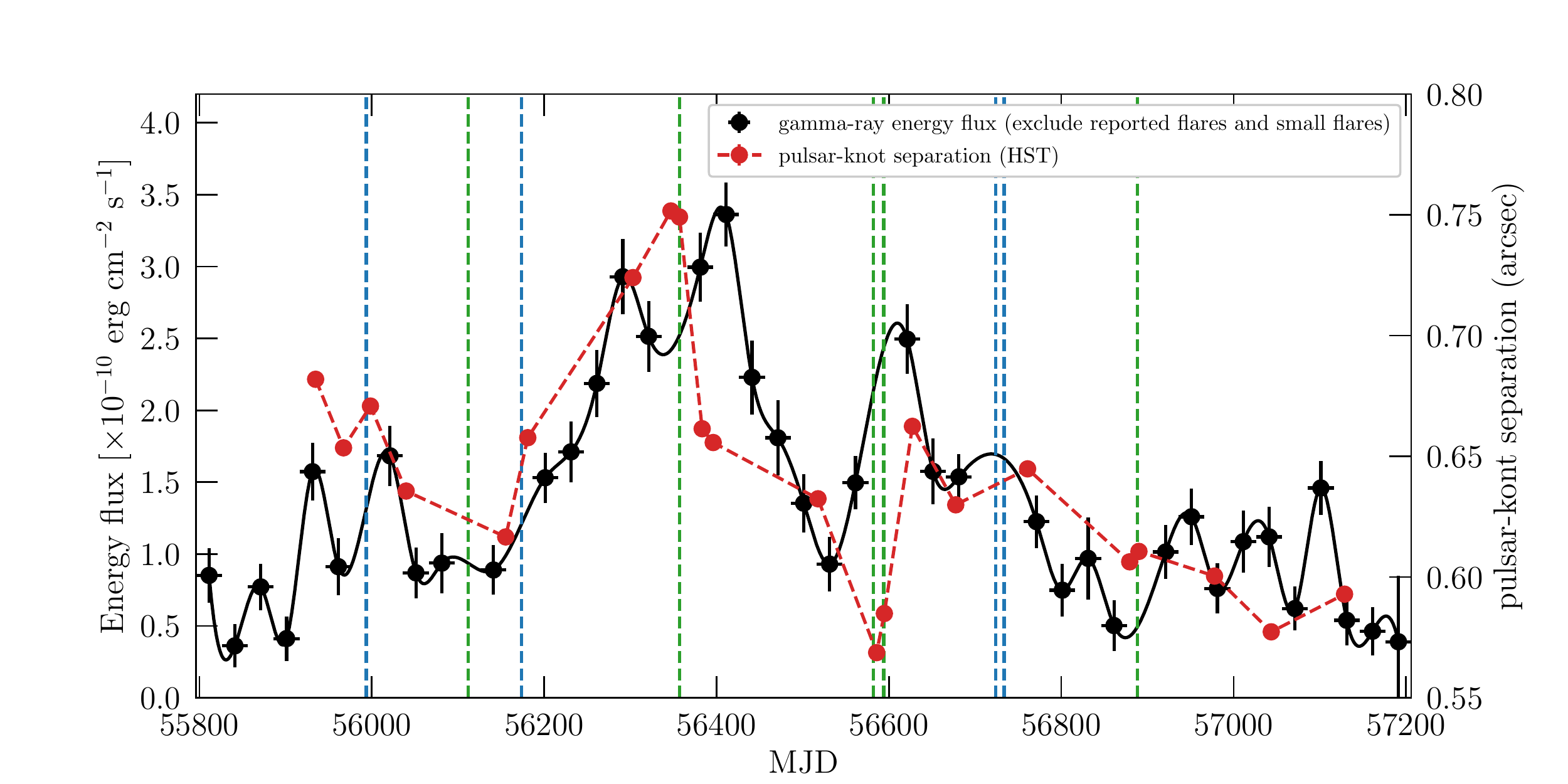}
\caption{30-day binned gamma-ray flux light curve (100 MeV - 500 GeV) of the Crab synchrotron component and the separation distance between the Crab pulsar and the inner knot.
Black data points: gamma-ray energy fluxes.
Red data points: separation distances between the Crab pulsar and the inner knot, measured by the {\textit {Hubble Space Telescope}}  \citep{Rudy2015}.
The small flares are shown by the blue lines, and the green lines show the reported flares.
}
\label{flux_knot}
\end{center}
\end{figure*}

\begin{figure}[tb]
\begin{center}
\includegraphics[width=100mm]{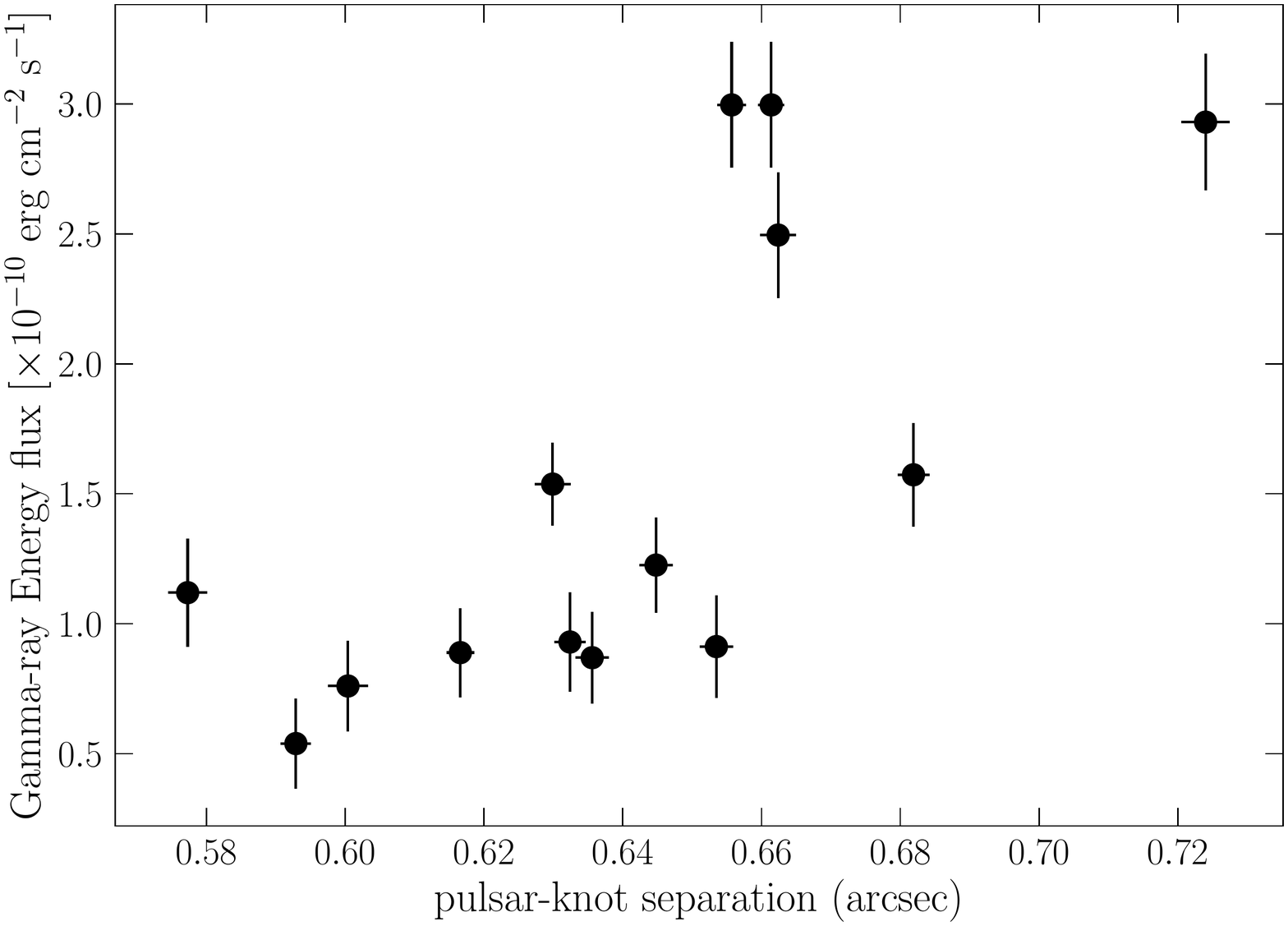}
\caption{Gamma-ray energy flux (100 MeV -- 500 GeV) vs. knot-pulsar separation.
The gamma-ray fluxes were averaged with  30-day bins and the knot-pulsar separation was observed by {\textit {HST}} within each 30-day bin.
The data for the knot-pulsar separation are from \cite{Rudy2015}.}
\label{corr_gamma_knot}
\end{center}
\end{figure}

To study a possible correlation between the month-scale variation of the Crab synchrotron emission and the knot-pulsar separation, we made a 30-day binned gamma-ray flux LC of the Crab synchrotron component through off-pulse analysis.
Off-pulse analysis of the Crab pulsar is necessary to investigate the month-scale variability because the variability amplitude is rather small and the variation can be hidden by the Crab pulsar emission.
The 30-day binned LC between MJD 55796 and 57206 is shown in Figure \ref{flux_knot} (black points and left axis).
The analysis procedure is the same as for the 5-day binned LC shown in Figure \ref{lc_5day};
the Crab PWN is modeled by Eq. (\ref{crab_stationary}) and the normalization, photon index ($N\sub{0,SYN}$ and $\Gamma\sub{0,SYN}$) of the Crab synchrotron component and the normalization of the isotropic diffuse emission are set free while the other sources are fixed to the baseline values.
We excluded data points that overlapped with any of the flares,
because we focus on the smaller flux variations in the longer-time-scale data.
Data points are interpolated by a cubic spline (black solid line).
The knot-pulsar separation observed by {\textit {Hubble Space Telescope}} ({\textit {HST}})  \citep{Rudy2015}\footnote{We choose the {\textit {Hubble Space Telescope}} data because the simultaneous Keck data may have unaccounted systematic errors~ \citep{Rudy2015}.} is shown in Figure \ref{flux_knot} as red points (scaled to the right axis).
Figure~\ref{corr_gamma_knot} compares the knot separations from the {\textit HST} observations with the corresponding gamma-ray energy fluxes based on the 30-day binned LC of Figure~\ref{flux_knot}.
The gamma-ray fluxes are visibly higher when the knot separation is larger.
Although this comparison is highly suggestive, the observations of the inner knot are relatively sparse compared to the various time scales of the gamma-ray flux. More evenly sampled observations by optical telescopes would be necessary to establish a firm correlation between the gamma-ray flux and the knot-pulsar separation.

%%%%%%%%%%%%%%%%%%%%%%%%%%%%%%%%%%%%%%%%%%%%%%%%%%%%%%%%%%%%%%%%%%%%%%%%%%%%%%%%%%%%%%%%%%%%%%%%%%%%%%%%
%%%%%%%%%%%%%%%%%%%%%%%%%%%%%%%%%%%%%%%%%%%%%%%%%%%%%%%%%%%%%%%%%%%%%%%%%%%%%%%%%%%%%%%%%%%%%%%%%%%%%%%%

%\clearpage
\section{Discussion}\label{section3}
\subsection{Flare statistics}
In Sec.~\ref{day_scale_ana}, we obtained the fluences, characteristic time scales, and photon indices for all 17 flares, including both the ``reported flares'' and the ``small flares''.
The relatively large number of detections allows us to discuss the features of the flares on a statistical basis.

As seen in Figure \ref{lc_corr}-(b), many of ``small flares'' have fluences smaller than
$2\times10^{-4}~\mathrm{erg\,cm^{-2}}$.  This result suggests that even weaker flares may exist, although it is
difficult to resolve such flares with  current instruments.  Superposition of unresolved weak flares may contribute to
the underlying variability of the LC seen in Figure \ref{lc_5day}.  Because of the fast variability the flares should be produced
through the synchrotron channel, and only electrons with the highest attainable energy may provide a significant
contribution in the GeV energy band.  If the spectral analysis of a flare
allows us to define the cut-off energy, we can use the single particle spectrum to define the energy
of the emitting particles, \(E_e\). The single-particle synchrotron spectrum is described by the following approximate
expression:
\begin{equation}\label{asymptotic_form_0}
  F_{\rm sp}(\omega) \propto \left(\frac\omega{\omega_c}\right)^{1/3}\exp\left(-\frac\omega{\omega_c}\right) \,.
\end{equation}
The critical frequency is defined as $\omega_c \equiv (3E^2_eeB)/(2m_e^3c^5)$, where
$E_e$, $e$, $m_e$, $c$, and $B$ are electron energy, the elementary charge, the electron mass, the speed of light, and
the magnetic field strength \citep{Rybicki}.

For the typical spectra revealed at GeV energies, detection of a cut-off in the {\it Fermi}-LAT band implies that emitting particles are accelerated by some very efficient acceleration mechanism or produced in a relativistically moving plasma.
Presently, it is broadly accepted that the magnetic field reconnection scenario provides the most comprehensive approach for the interpretation of the flaring emission \cite[see, e.g.,][]{2012ApJ...754L..33C,flare_review}. We therefore present some basic estimates in the framework of this scenario  \cite[see][for a more detailed consideration]{2005MNRAS.358..113L}. We assume that the reconnection proceeds in the relativistic regime, and  we ignore some processes, e.g., the impact of the guiding magnetic field,  which potentially might be very important. A detailed consideration of magnetic reconnection is beyond the scope of this paper.

If we assume the flares originate from magnetic reconnection events, then some key aspects of the flare emission are determined by the geometry of the reconnection region.
The maximum energy that an electron can gain in the reconnection region can be estimated as
\begin{equation}
    E\sub{e, max} = eB_0L \,.
\end{equation}
Here $B_0$ and $L$ are the strength of the reconnecting magnetic field and the length of the layer, respectively. It was assumed here that
the electric field at the reconnection region is equal to the initial magnetic field.
The length of the reconnection region, \(L\), determines the flare rise time, \(\tau\sub{rise}\simeq L/c\).
The total magnetic energy that is dissipated in the region is given by
\begin{equation}
    W_B = \frac{cB\sub{0}^2}{4\pi}aL\frac{L}{c} = \frac{a_L L^3B\sub{0}^2}{4\pi}\,,
\end{equation}
where $a=a_L L$ represents the width of the reconnection region (i.e., its length in the direction perpendicular to the electric field).
The emission produced by particles accelerated beyond the ideal MHD limit might be highly anisotropic  \citep{2007ApJ...655..980D,2012ApJ...754L..33C},
as electrons may lose energy over just a fraction of the gyro radius and photons are emitted within a narrow beaming cone.
The beaming solid angle is determined by  $\Delta \Omega  =\mathrm{min}[ \pi(ct\sub{cool}/r\sub{g})^2, \,4\pi]\,$,
where $t\sub{cool}\propto E\sub{e}^{-1}B^{-2}$ is the synchrotron cooling time and $r\sub{g}\propto E\sub{e}B^{-1}$ is the gyro radius of the electron.
The values of the solid angle describe two distinct regimes: $\Delta\Omega=\pi(ct\sub{cool}/r\sub{g})^2$ corresponds to the strong beaming case whereas $\Delta\Omega=4\pi$ represents the isotropic radiation case.

The observed fluences can be written as \(F= \xi W_B/\left(D^2\Delta \Omega\right)\), which yields
\begin{equation}\label{beam}
    F\sub{beaming}  \propto B\sub{0}^{-5/2}(\hbar\omega_c)^{7/2}
\end{equation}
for the strong beaming case and
\begin{equation}\label{iso}
    F\sub{iso} \propto B\sub{0}^{-5/2}(\hbar\omega_c)^{3/2}
\end{equation}
for the isotropic case. Here $\xi$ and $D$ represent the radiation efficiency and the distance to the Crab \ac{pwn}, respectively.

\begin{figure}[btp]
\begin{center}
\includegraphics[width=100mm]{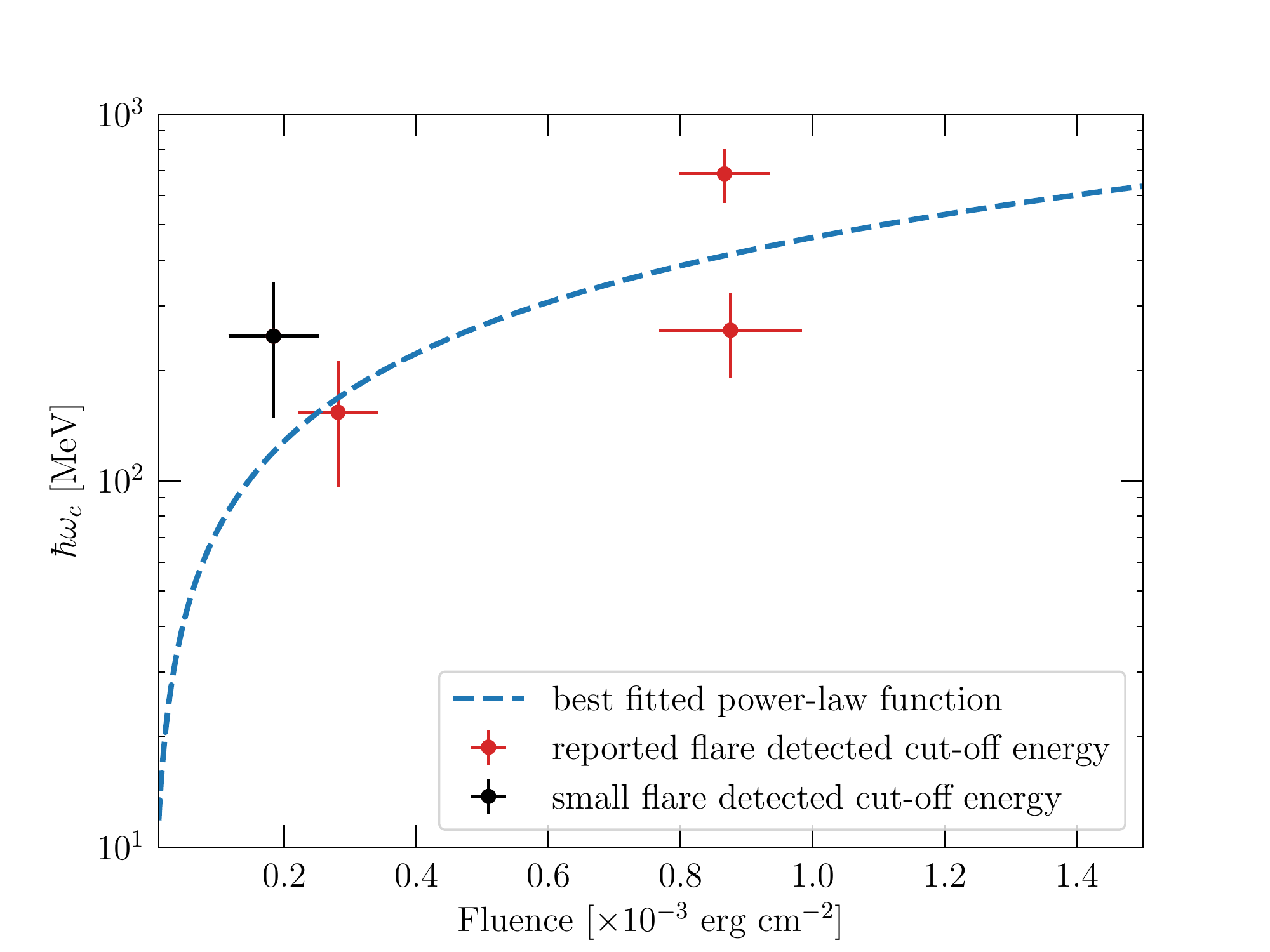}
\caption{Fluence vs. cut-off energy for each flare in which significant spectral cut-off was observed.
Black data point: small flare 7 (b) Red points:  2011 April flare, 2013 March flare, and 2013 October flare (a).
The critical synchrotron energy ($\hbar\omega_c$) is defined by the observed cut-off energy.
The dashed blue line shows the best fitted power-law function for those data points.
}
\label{fluence_omega}
\end{center}
\end{figure}

The fluence $F\sub{SYN}$ vs. the critical photon frequency is shown in Figure~\ref{fluence_omega}.
The dependence between these quantities was fitted with a power-law function:
\begin{equation}\label{pl}
    \hbar \omega_c \propto F\sub{SYN}^{p}\,.
\end{equation}
We include here only the flares that allow the detection of the cut-off energies (i.e., small flare 7 (b), 2011 April flare, 2013 March flare and 2013 October flare (a))\footnote{We do not consider 2013 October flare (b) whose rise time is not determined.}.
For these data we obtained \(p=0.74\pm0.32\) as the best-fit value.
The best-fit approximation is shown in Figure \ref{fluence_omega} as a blue dashed line.

The dependence of the fluence on the critical synchrotron energy for the anisotropic regime, \(p=0.286\), does not seem to agree with the revealed best-fit approximation, unless the magnetic field changes strongly between the flares. However, the isotropic regime, \(p=0.66\), might be consistent with the dependence for same strength of the field, although the error bars are significant.
This might be taken as a hint for the reconnection origin of the variable emission, but more detailed consideration makes this possibility less feasible. In particular, Eq. (\ref{iso}) allows us to obtain estimates for the strength of the magnetic field as
\begin{equation}\label{iso_B}
    B\sub{iso} \simeq 400 ~a_L^{\frac{2}{5}}\xi^{\frac{2}{5}}\left(\frac{D}{2~\mathrm{kpc}}\right)^{-\frac{4}{5}}\left(\frac{\hbar\omega\sub{c}}{300~\mathrm{MeV}}\right)^{\frac{3}{5}} \left(\frac{F\sub{SYN}}{6\times10^{-4}~\mathrm{erg~cm^{-2}}}\right)^{-\frac{2}{5}}\,\mathrm{\upmu G}\,,
\end{equation}
As shown below, the detected rise and decay time scales require a significantly stronger magnetic field.
This discrepancy implies that the magnetic field reconnection cannot be readily taken as the ultimate explanation for the origin of the variable GeV emission detected from the Crab \ac{pwn}, suggesting that other phenomena, e.g., Doppler boosting, plays an important role \cite[see, e.g.,][]{Rolf2012}.\\
%%%%%%%

Reconstruction of the spectrum of the emitting particles for the flares that do not show a well-defined cut-off energy is less straightforward.
Let us assume that the electron spectrum is described by a power-law with an exponential cut-off: \(dN_e/dE_e\propto E^{-\alpha} {\exp}\left(-E_e/E_{\rm max}\right)\). The modeling of the emission from the Crab PWN  implies that the high-energy part of the electron spectrum is characterized by \(\alpha\simeq3.23\) and the cut-off energy is in the PeV band \(E_{\rm max}\simeq3\rm \,PeV\) \citep[as obtained by][in the framework of the constant B-field model]{Meyer2010}.
The total synchrotron spectrum is then
\begin{equation}\label{asymptotic_form_1}
  F(\omega) \propto \int dE_e' E'_e{}^{-\alpha} {\exp}\left(-E'_e/E_{\rm max}\right)\left(\frac\omega{\omega_c}\right)^{1/3}\exp\left(-\frac\omega{\omega_c}\right) \,.
\end{equation}
For the spectra, which do not allow determination of the cut-off frequency, one should expect \(\hbar\omega_c(E_{\rm max})\lesssim 100\rm\,MeV\) or  \(\hbar\omega_c(E_{\rm max})\gg100\rm\,MeV\), the later possibility is however robustly excluded by the synchrotron radiation theory and the broadband spectrum measured with {{\it Fermi}}--LAT \footnote{The observed cut-off energy between {\it{COMPTEL}} and {{\it Fermi}}--LAT band was $97 \pm 12$ MeV \citep{Crab_pulsar2010}}.
In the regime \(\omega \geq \omega_c(E_{\rm max})\), the above integral can be computed with the {\it steepest descent} method yielding
\begin{equation}\label{asymptotic_form}
  F(\omega) \propto E_{\rm max}^{-\alpha+1}\left(\frac{\omega}{\omega_c(E_{\rm max})}\right)^{-\frac{6\alpha-5}{18}}\exp\left(-\frac3{2^{2/3}}\left(\frac\omega{\omega_c(E_{\rm max})}\right)^{1/3}\right) \,,
\end{equation}
where the dependence on \(E_{\rm max}\) and \(B\) is kept.
The photon index at \(\omega\) can be obtained as
\begin{equation}
    \Gamma\sub{SYN} \simeq 1-\frac{\frac{dF(\omega/\omega_c(E_{\rm max}))}{d\omega}}{F(\omega/\omega_c(E_{\rm max}))}\omega \,,
\end{equation}
which can be derived analytically: %This yields the following approximate relation between the photon index and the critical frequency:
\begin{equation}\label{omega_phindex}
    \Gamma\sub{SYN}=\frac{6\alpha+13}{18}+\frac1{2^{\nicefrac23}}\left(\frac{\omega}{\omega_c(E_{\rm max})}\right)^{\nicefrac13}\,.
\end{equation}

The variable gamma-ray emission might originate from the distribution of non-thermal electrons that are responsible for the broad-band emission from the Crab PWN, for example if the strength of the magnetic field fluctuates or the electron cut-off energy changes. To probe these possibilities we set \(\alpha=3.23\) and study the relation between the flare flux and photon index. If the variation is caused by a change of the magnetic field (i.e., \(E_{\rm max}=\rm const\)), then
\begin{equation}\label{eq:eflux_gamma_1}
F(\omega)\propto \left(\Gamma\sub{SYN}-1.8\right)^{-2.4}\exp\left(-3\left(\Gamma\sub{SYN}-1.8\right)\right)\,.
\end{equation}
If the variability of the gamma-ray emission is caused by a change of the electron cut-off energy, then there is an additional factor in the expression that determines the flux level:
\begin{equation}\label{eq:eflux_gamma_2}
F(\omega)\propto \left(\Gamma\sub{SYN}-1.8\right)^{0.95}\exp\left(-3\left(\Gamma\sub{SYN}-1.8\right)\right)\,.
\end{equation}

\begin{figure}[tbp]
\begin{center}
\includegraphics[width=100mm]{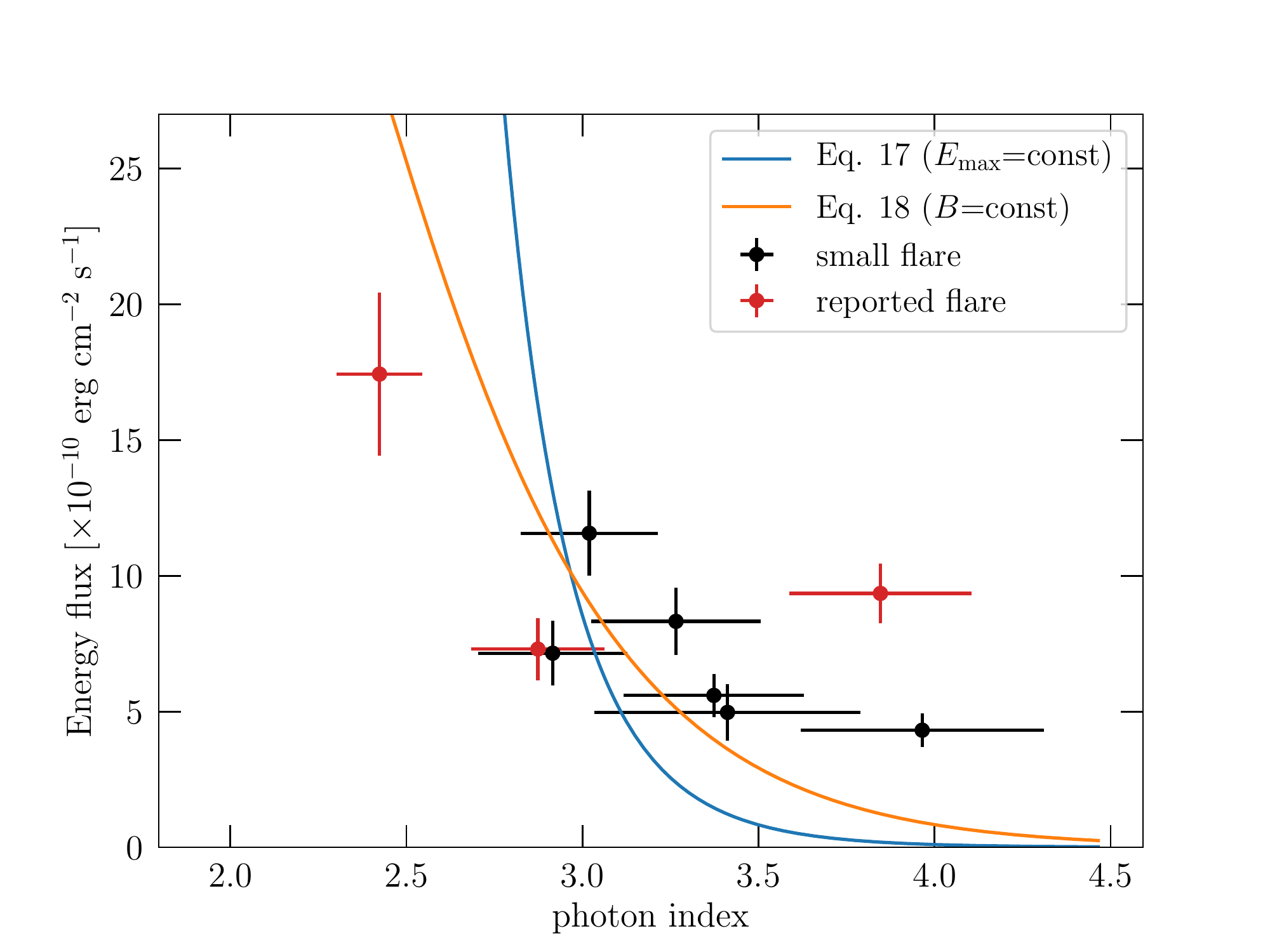}
\caption{
Energy flux vs. photon index for the Crab flares that resulted in non-detection of the cut-off energy.
Black data points: ``small flares". Red data points: ``reported flares".
Two lines show the best-fit approximation of the data points by Eq.~(\ref{eq:eflux_gamma_1}) ($E_{\rm max} =$const.: blue line) and Eq.~(\ref{eq:eflux_gamma_2}) ($B=$const: orange line).
}
\label{eflux_gamma}
\end{center}
\end{figure}
The plot of the energy flux and photon index for the ``small flares" and the ``reported flares" is shown in Figure~\ref{eflux_gamma}.
From the figure it can be seen that the data points seem to be inconsistent with Eq.~(\ref{eq:eflux_gamma_1}), which corresponds to the case when
the flare is generated by a changing magnetic field. Although Eq.~(\ref{eq:eflux_gamma_2}) agrees better with the data,
the discrepancy is still very significant. This suggests that a  one-zone model is inadequate for the study of the observed variability, and a more detailed model is required. In particular, the properties at the flare production site might be different from the typical conditions inferred for the nebula, implying the need for a multi-zone configuration. In what follows we try to infer the conditions at the flare production site using a simple synchrotron model.
%%%%%%%
\subsection{The origin of small flares}
All ``small flares" except for one episode allow defining the rise and decay time-scales based on the 1.5-day binned light curves. Here, we discuss constraints on the magnetic field strength assuming a synchrotron origin of the emission.
For the sake of simplicity we adopt a model that assumes a homogeneous magnetic field in the flare production site, which does not move relativistically with respect to the observer. Thus the rise time is associated with the particle acceleration time-scale:
\begin{equation}
    \tau\sub{acc} \simeq 10\eta\left(\frac{E\sub{e}}{1\rm\,PeV}\right)\left(\frac{B}{100\rm\,\upmu G}\right)^{-1} \rm \, days\,,
\end{equation}
where \(\eta\), $E\sub{e}$ and $B$ are acceleration efficiency (\(\eta\rightarrow1\) for particle acceleration by reconnection), electron energy and magnetic field strength, respectively.
% ;  \(E={\cal E}\rm\,PeV\)  and \(B=100{\cal B}\rm\,\upmu G\) are electron energy and the magnetic field strength, respectively.
Since variability is observed for \(\sim100\rm \, MeV\) emission, we obtain
\begin{equation}
    E\sub{e}\simeq 4 \left(\frac{B}{100\rm\,\upmu G}\right)^{-1/2}\left(\frac{\ve}{100\rm\,MeV}\right)^{1/2}\rm\,PeV\,,
  %{\cal E}= 4 {\cal B}^{-1/2} \ve ^{1/2}\,.
\end{equation}
where $\ve$ is synchrotron photon energy.
This yields an acceleration time of
\begin{equation}\label{eq:acc}
    %\tau\sub{acc} =40\eta^{2/3} {\cal B}^{-3/2}\ve ^{1/2} \,,
    \tau\sub{acc} \simeq 40\eta \left(\frac{B}{100\rm\,\upmu G}\right)^{-3/2}\left(\frac{\ve}{100\rm\,MeV}\right)^{1/2} \rm\,days\,,
\end{equation}
which translates to the following limitation on the magnetic field strength:
\begin{equation}\label{eq:acc_b}
    %{\cal B} >10\eta^{2/3} \tau\sub{acc}^{-2/3}\ve ^{1/3} \,.
    B \simeq 1\eta^{2/3} \left(\frac{\tau\sub{acc}}{1\rm\, day}\right)^{-2/3}\left(\frac{\ve}{100\rm\,MeV}\right)^{1/3}\rm\, mG\,.
\end{equation}
The observational requirement of \(\tau\sub{acc}\approx1\) day implies an extremely strong magnetic field or an acceleration with efficiency exceeding the ideal \ac{mhd} limit: \(\eta<1\).
The cooling time, however, which is associated with the decay time (\(\approx2\) days), does not depend on the acceleration efficiency and allows us to obtain the following constraint for the magnetic field
\begin{equation}\label{eq:cool}
  B \simeq 1\left(\frac{\tau\sub{cool}}{2\rm\,days}\right)^{-2/3}\left(\frac{\ve}{100\rm\,MeV}\right)^{-1/3} \rm\, mG\,.
\end{equation}
This estimate is valid for the isotropic radiation regime, which, in particular, implies negligible Doppler boosting and homogeneity of the magnetic field. These assumptions are most likely  violated for the strong flares \cite[see, e.g.,][]{Rolf2012}. We nevertheless adopt these assumptions here as we aim to test the consistency of the ``small flares'' with these assumptions {\it ex adverso}.

Given that (i) both the rise and decay time-scales are short, and (ii) the dependence on the photon energy in Eqs.~(\ref{eq:acc}) and (\ref{eq:cool}) is inverse,  the magnetic field in the production region needs to be very strong, approaching the \(\rm mG\) value. This estimate seems to be inconsistent with Eq.~(\ref{iso_B}), posing certain difficulties for the reconnection scenario.

In \acp{pwn} magnetic fields can provide an important contribution to the local pressure:
\begin{equation}
  P\sub{B}=\frac{B^2}{8\pi}=4\times10^{-10} \left(\frac{B}{100\rm\,\upmu G}\right)^{2}\rm\,erg\,cm^{-3}\,.
\end{equation}
On the other hand, \acp{pwn} are expected to be nearly isobaric systems; thus the characteristic pressure in the Crab \ac{pwn} can be determined by the location of the pulsar wind termination shock:
\begin{equation}
  P\sub{PWN} \simeq \frac{2}{3}\times\frac{\dot{E}\sub{SD}}{4\pi R\sub{TS}^2c} \simeq 5\times 10^{-9}\left(\frac{\dot{E}\sub{SD}}{5\times10^{38}~\mathrm{erg~s^{-1}}}\right)\left(\frac{R\sub{TS}}{0.14~\mathrm{pc}}\right)^{-2}~\mathrm{erg\,cm^{-3}}\,,
\end{equation}
where $\dot{E}\sub{SD}$ and $R\sub{TS}$ are the spin-down power of the central pulsar and the location 1of the termination shock, respectively.
Such total pressure implies the magnetic field of $B\lesssim400\rm\,\upmu G$, which appears to be significantly weaker than the strength required for acceleration and cooling of the particles responsible for the ``small flares.''
Strong local variations of the pressure/magnetic field at the termination shock seen in the 3D MHD simulations \citep{2014MNRAS.438..278P, 2016JPlPh..82f6301O} may mitigate the discrepancy between the estimated magnetic field and the one required to accelerate and cool the particles responsible for the ``small flares."

This result implies that the production of small flares under conditions typical for the nebula is difficult. One of the common approaches to relax the constraints imposed by the fast variability is to involve a relativistically moving production site \citep{Komissarov2011,Lyutikov2016}.
In this case, the apparent variability time-scales are shorted:
\begin{equation}
  \tau=\frac{\tau'}\delta\,,
\end{equation}
where \(\tau'\) and \(\delta\) are the plasma co-moving frame time scale and the Doppler boosting factor, respectively.
%\begin{equation}\label{doppler}
%    \delta \equiv\frac{1}{\gamma(1-\beta\cos{\theta})}\,.
%  \end{equation}
%Here $\gamma$ is the fluid bulk Lorentz factor in the downstream and $\theta$ is the angle between the line of sight and the velocity vector of the fluid bulk motion.
In addition, the emission frequency changes as well, so the co-moving frame photon energy is \(\ve'=\ve/\delta\). Thus, one obtains that the requirements for magnetic field strength, Eqs.~(\ref{eq:acc_b}) and (\ref{eq:cool}) are relaxed by a factor  \(\delta\) and \(\delta^\frac13\), respectively.
The maximum magnetic field strength consistent with the nebular pressure, $B=400\rm\,\upmu G$, gives the following lower limit for the bulk Lorentz factors from Eqs.~(\ref{eq:acc_b}) and (\ref{eq:cool}):
\begin{equation}
    %\delta > 10\eta^{2/3}\tau\sub{acc}^{-2/3}\ve ^{1/3}{\cal B}^{-1}
    \delta > 3\eta^{2/3}\left(\frac{\tau\sub{acc}}{1\rm\,day}\right)^{-2/3}\left(\frac{\ve}{100\rm\,MeV}\right)^{1/3}\left(\frac{B}{400\rm\,\upmu G}\right)^{-1}
 %   \delta > \eta^{2/3}\tau\sub{acc}^{-2/3}{\ve}^{1/3}B^{-1}
\end{equation}
\begin{equation}
    \delta > 30\left(\frac{\tau\sub{cool}}{2\rm\, days}\right)^{-2}\left(\frac{\ve}{100\rm\,MeV}\right)^{-1}\left(\frac{B}{400\rm\,\upmu G}\right)^{-3}\rm .
%   \delta > \tau\sub{cool}^{-2}{\ve}^{-1}B^{-3}
\end{equation}
The existence of such high bulk Lorentz factors in the termination shock downstream region seems to be challenging, but, probably cannot be excluded from first principles.    For example,  the bulk Lorentz factor of a weakly magnetized flow passing through an inclined relativistic shock depends only on the inclination angle \(\alpha\): \(\gamma= 3/(\sqrt{8}\sin\alpha)\)  \citep[see, e.g.,][]{2002AstL...28..373B}. Thus, the required bulk Lorentz factor, \(\gamma\simeq15\), can be achieved if the pulsar wind velocity makes an angle of \(\alpha\sim5^\circ\) with the termination shock.
Detailed \ac{mhd} simulations are required to verify the feasibility of producing variable GeV emission in a relativistically moving plasma.
Such MHD simulations should allow a realistic magnetic field and plasma internal energy in the flow that is formed at the inclined termination shock. These parameters are required to define the spectrum of non-thermal particles in the flow and in turn to compute the synchrotron radiation \citep{2014MNRAS.438..278P,Lyutikov2016}.

As discussed in the literature  \citep[see,
e.g.,][]{Komissarov2011}, the ``inner knot" may be associated with the part of the post-shock flow that is characterized by
a large Doppler factor, and a significant fraction of \(100\rm\,MeV\) gamma rays might be produced in this region.
Some spectral features seen in the hard X-ray band can by interpreted as hints supporting this hypothesis.
The spectral energy distribution of the non-thermal emission and X-ray morphology revealed with the {\textit {Chandra X-ray Observatory}} are consistent
with \ac{mhd} models that assume that the non-thermal particles are accelerated at the pulsar wind termination shock
and are advected with the nebular \ac{mhd} flow.
However, there is no observational evidence indicating that the PeV particles that are responsible for the synchrotron gamma-ray emission belong to the same distribution.
The COMPTEL data suggest a spectral hump around $\sim$ 1 MeV  \citep{Crab_comptel}.
In addition, \textit{NuStar} found a spectral break with $\Delta \Gamma\sim0.25$ at $\sim 9$ keV in the tourus region \citep{Nustar}, and SPI on {\textit{INTEGRAL}} also detected a spectral steeping above 150 keV  \citep{Crab_SPI}.
These observational results may imply the existence of another spectral component above $\sim 10$ keV that is produced by a different population of non-thermal particles than the optical and soft X-ray emission  \citep{Aharonian1998,2019MNRAS.489.2403L}.

To study the possible relation of the \(100\rm\,MeV\) gamma-ray emission to the inner knot, we constructed the 30-days-binned gamma-ray LC of the Crab PWN synchorotron component, i.e., nebula emission with excluded flares and
``small flares''.  We found a naive correlation between this LC and the pulsar to inner-knot separation, as shown in
Figure~\ref{flux_knot}. This tendency is, however, inconsistent with the prediction of the \ac{mhd} model of
 \citet{Komissarov2011}.  Further observations of the ``inner knot'' can test this relation.
%%%%%%%%%%%%%%%%%%%%%%%%%%%%%%%%%%%%%%%%%%
%%%%%%%%%%%%%%%%%%%%%%%%%%%%%%%%%%%%%%%%%%
\\
\section{Conclusions}\label{section4}
We performed a systematic search for gamma-ray flares from the Crab PWN using 7.4 years of data from the
\textit{Fermi} LAT.  Our analysis used the off-pulse window of the Crab pulsar.  In addition to the flares
reported in the literature (``reported flares''), {\SF} lower-intensity flares (``small flares'') are found in this work.
The synchrotron component of the Crab PWN shows highly variable emission,  indicatings that the ``small flares'' also originate from synchrotron radiation.
The “small flares” are different from the stationary synchrotron component not only by larger fluxes but also by harder photon indices.
We determined the characteristic scales of both rise and decay times for the ``small flares'' based on 1.5-day
light curves, and we determined that the ``small flares'' are characterized by a day-scale variability.
Apart from two exceptionally bright flares in 2011 April and 2013 March, the ``small flares'' and the ``reported
flares'' show the same range of photon index with a harder-when-brighter trend. We tested the distribution of the flare
parameters against predictions from a simple reconnection model. Although the dependence of the emission fluence on the
observed cut-off energy appeared to be consistent with the model prediction based on the isotropic emission region for some flares, the implied magnetic
field appeared to be too weak to reproduce the observed variability. In order to explain the short time
variability, a strong magnetic field, $\approx1~\mathrm{mG}$, is required.  Such a high magnetic field at the production
site of the ``small flares'' implies a magnetic pressure significantly exceeding the value that is anticipated from the
position of the pulsar wind termination shock.  This challenges the conventional view on the origin of the
\(100\rm\,MeV\) gamma-ray emission from the Crab PWN.
The requirement of the strong magnetic field can be relaxed by assuming that the production site of the ``small flares'' moves relativistically with respect to the observer.
In this case, rather high Doppler boosting factors, \(\delta\gtrsim10\), are required.
Such Doppler factors are attainable in the termination shock downstream if the pulsar wind passes through an inclined shock, which makes an angle of \(\alpha\sim5^\circ\) with the wind velocity.
The magnetic field and plasma internal energy in the flow also depend on $\alpha$; thus more detailed \ac{mhd} simulations are required to verify the possibility of the production of the ``small flares'' by relativistically moving plasma.
\acknowledgments
{The \textit{Fermi} LAT Collaboration acknowledges generous ongoing support
from a number of agencies and institutes that have supported both the
development and the operation of the LAT as well as scientific data analysis.
These include the National Aeronautics and Space Administration and the
Department of Energy in the United States, the Commissariat \`a l'Energie Atomique
and the Centre National de la Recherche Scientifique / Institut National de Physique
Nucl\'eaire et de Physique des Particules in France, the Agenzia Spaziale Italiana
and the Istituto Nazionale di Fisica Nucleare in Italy, the Ministry of Education,
Culture, Sports, Science and Technology (MEXT), High Energy Accelerator Research
Organization (KEK) and Japan Aerospace Exploration Agency (JAXA) in Japan, and
the K.~A.~Wallenberg Foundation, the Swedish Research Council and the
Swedish National Space Board in Sweden.

Additional support for science analysis during the operations phase is gratefully
acknowledged from the Istituto Nazionale di Astrofisica in Italy and the Centre
National d'\'Etudes Spatiales in France. This work performed in part under DOE
Contract DE-AC02-76SF00515.}

M.A. is supported by the RIKEN Junior Research Associate Program.
This work was supported by KAKENHI Grant Numbers 18H03722 and 18H05463 (Y.U.).
\bibliography{refrence}

\end{document}